\documentclass[showpacs, twocolumn, preprintnumbers,amsmath,amssymb]{revtex4}
\usepackage{mathrsfs}
\usepackage{graphicx}
\usepackage{subfigure}
\usepackage{indentfirst}
\usepackage{amsfonts}
\usepackage{amssymb}%
\usepackage{ulem}
\usepackage{color}

\def\al{\alpha}
\def\be{\beta}
\def\ga{\gamma}
\def\de{\delta}
\def\ga{\gamma}
\def\si{\sigma}
\def\Ga{\Gamma}
\def\prt{\partial}
\def\ch{\mathrm{ch}}
\def\sh{\mathrm{sh}}

\newcommand{\bea}{\begin{eqnarray}}
\newcommand{\eea}{\end{eqnarray}}
\newcommand{\bit}{\begin{itemize}}
\newcommand{\eit}{\end{itemize}}
\newcommand{\hf}{\frac{1}{2}}
\newcommand{\nn}{\nonumber\\}
\newcommand{\ie}{{\it i.e.}}
\newcommand{\eg}{{\it e.g.}}
\newcommand{\etl}{{\it et.al.}}

\providecommand{\Journal}[4] {#1 {\bf#2}, #4 (#3)}
\providecommand{\IJMPB}{Int. J. Mod. Phys. B} %
\providecommand{\CQG}{Class. Quantum Grav.}
\providecommand{\PRL}{Phys. Rev. Lett.} %
\providecommand{\PR}{Phys. Rev.} %
\providecommand{\PRA}{Phys. Rev. A.} %
\providecommand{\PRP}{Phys. Rep.} %
\providecommand{\NT}{Nature} %
\providecommand{\NP}{Nature Phys.} %
\providecommand{\OpEx}{Opt. Express} %
\providecommand{\AJP}{Am. J. Phys.} %
\providecommand{\JMP}{J. Math. Phys.} %
\providecommand{\JMO}{J. Mod. Opt.} %
\providecommand{\JAP}{J. Appl. Phys.} %
\providecommand{\JETP}{J. Exptl. Theoret. Phys.}%
\providecommand{\JPA}{J. Phys. A} %
\providecommand{\JPC}{J. Phys. C} %
\providecommand{\JPGNPP}{J. Phys. G: Nucl. Part. Phys.} %
\providecommand{\JRNIST}{J. Res. Natl. Inst. Stand. Technol.}%
\providecommand{\RMP}{Rev. Mod. Phys.} %
\providecommand{\RSI}{Rev. Sci. Instrum.} %
\providecommand{\SE}{Science} %
\providecommand{\SSC}{Solid State Commun.} %
\providecommand{\SJNP}{Sov. J. Nucl. Phys.}

\begin{document}
\title{\Large {The neutron returning time in a linear potential}}
\author{Zhi Xiao}
\email{spacecraft@pku.edu.cn}
\affiliation{Department of Mathematics and Physics, North China Electric Power University, Beijing 102206, China}
\author{Shuang Zheng}
\affiliation{Department of Mathematics and Physics, North China Electric Power University, Beijing 102206, China}
\author{Ji-Cai Liu}
\email{jicailiu@ncepu.edu.cn}
\affiliation{Department of Mathematics and Physics, North China Electric Power University, Beijing 102206, China}

\begin{abstract}
In this paper, we calculate the quantum time delays for neutron scattering off the Earth's linear gravitational potential.
The quantum time delays are obtained by subtracting the classical returning time (CRT) from the
Wigner time, the dwell time and the redefined Larmor time respectively.
Different from the conventional definition, our Larmor time is defined by aligning the magnetic field along
the neutron propagation direction, and this definition does give reasonable results for motions through a free region and a square barrier.
It is worth noting that in the zero magnetic field limit, the Larmor time coincides well with the CRT,
which is due to the special shape of linear barrier, and may have some relevance to the weak equivalence principle.
It is also found that the classical forbidden region plays an essential role for the dwell time $\tau_{_\mathrm{DW}}$
to match with the CRT, and the difference between the dwell and the phase times, \ie, the self-interference time delay, is barrier shape sensitive and clearly shows the peculiarity of the linear barrier.
All the time delays are on the order of sub-millisecond and exhibit oscillating behaviors,
signaling the self-interference of the scattering neutron, and the oscillations become evident only when the de Broglie wavelength $\lambda_k=2\pi/k$ is comparable to the characteristic length $L_c=[2m^2g/\hbar^2]^{-1/3}$.
If the time delay measurement is experimentally realizable,
it can probe the quantum nature for particle scattering off the gravitational potential in the temporal domain.
\end{abstract}

\maketitle
\section{introduction}
Quantum tunneling is a pure quantum phenomenon without any classical analogy, and the relevant traversal time
in tunneling is an intensively debated issue since the early days of quantum mechanics \cite{EDQT}.
Many controversial definitions on tunneling time coexist along the debate.
To name a few, the Wigner phase time \cite{EPW}\cite{Bohm}, dwell time \cite{Smith} and Larmor time \cite{BRLarmor}
are among the most widely discussed tunneling times,
and they capture distinctive features of quantum tunneling by definition \cite{Yamada2004}.
Aside from the lack of a unified definition of tunneling time, not all time definitions essentially describe traversal time \cite{BITLM}\cite{Winful2006},
and hence the interesting superluminal {\it Hartman effect} \cite{Hartman} is only an artifact of the misinterpretation \cite{Winful2006}\cite{ZXRT2015}.
Nevertheless, a unified derivation of Larmor time, B$\ddot{\mathrm{u}}$ttiker-Landauer time \cite{Buttiker1982},
Wigner phase time and Pollak-Miller time \cite{PMT}, in terms of the Gell-Mann--Hartle decoherence functionals has been
obtained recently \cite{Yamada2004}, where various times have been neatly classified as the total time
for particle sojourning in the barrier and the ultimate time difference for a particle traversal across the barrier.
One can consult to several excellent review articles for further reading \cite{HSRMP1}\cite{BITLM}\cite{Winful2006}\cite{ASTT2015}.

In our opinion, there are at least two reasons for the issue being suspended more than 80 years.
First, it is only very recently one seems to reach an acceptable answer to the question of
{\it whether traversal time is best viewed as a distribution, or as a single time scale} \cite{BITLM}.
From the path integral approach, Ref. \cite{Yamada1999} undeniably pointed out that
at least for opaque barriers, there is no room for a definable unique tunneling time.
Further, tunneling time does not possess a direct probability distribution, rather it can be assigned a
somewhat vague distribution,
a distribution where temporal interference occurs between different times corresponding to ``distinctive classical paths"
(in the sense of path integral) for a particle traveling beneath the barrier,
just like spatial interference in a double-slit experiment for a particle traveling through alternative slits \cite{Yamada1999}.
Besides, for different definitions of tunneling times, the exact meanings of temporal distributions maybe quite distinct
\cite{dwellDis}\cite{NelsonQMDis}.
Second, the estimated time scales for most systems are far too small.
For example, the possible ionization tunneling delay is about $10\sim100$ attoseconds
(1as=$10^{-18}$s, approximately the time light orbiting around a circle with Bohr radius),
so to pin down the time issue experimentally is extremely difficult.
However, the situation is dramatically changed with the advent of attoclock \cite{AttoClock},
whose time resolution can already reach the level of attosecond \cite{PhotoemiTD}\cite{ASAS2019}.
This ultra-precise temporal resolution revives a surge of interest in
tunneling time \cite{TTWM2016}\cite{TIT2016}\cite{ITTDS2017}\cite{IAMTT2015},
and also makes a satisfactory answer to the tunneling issue more urgent.
Recently, evidence of finite tunneling time comes from the studies of strong field ionization in multi-electron atoms \cite{EvidenceArKr},
and some theoretical calculations \cite{TTWM2016} also favor tunneling as a finite process,
while other studies claim supports for instantaneous tunneling \cite{IAMTT2015}\cite{ASAS2019}.
Up to now, there is still no theoretical consensus on whether tunneling is an instantaneous or a
finite process.

To shed new light on the issues mentioned above,
we utilize an ultracold neutron (UCN) scattering off a linear gravitational potential as an illustrative model.
Firstly, the linear potential provides an alternative analytical example to the study of tunneling time besides
the extensively discussed square barrier. This may be used to explore the barrier-shape sensitive time definitions
(\eg, dwell time \cite{TTWM2016}), which cannot be uncovered by the square barrier itself.
Secondly, due to the tiny kinematic energy, the characteristic quantum time delay is on the order of sub-millisecond
(will be shown in the following), which is much easier to calibrate than the ultrafast ionization tunneling mentioned above.
Actually, for slow-moving massive neutral particles such as atoms, the tunneling time measurements have successfully achieved
the microseconds accuracy \cite{AtomT}\cite{AtomOL}\cite{BECTT}.
Though UCN is neither easy to prepare nor to manipulate compared to the precisely controllable cold atoms with laser field,
it still allows precise measurement, such as the precisely measured transition frequencies between different gravitational states,
which has already attained $0.1$Hz accuracy \cite{GRS}.
So we believe that with sophisticated design, precise time measurement with UCN may still be possible.
Further, this simple model can also evade the complication due to multi-time scales in ionization tunneling,
where the temporal scales include not only the tunneling delay, but also the resonance lifetimes of bound states \cite{TTWM2016}.

Though for a linear potential, tunneling delay has already been discussed by Davies \cite{Davies}
aiming to validate the weak equivalence principle,
our study here is to demonstrate a potentially testable time delay caused by the linear barrier from
the discussion of phase time, dwell time and Larmor time.
Note our definition on Larmor time is experimentally feasible, as only weak magnetic field
instead of zero magnetic field limit \cite{BRLarmor} is required.
Moreover, different from the original definition \cite{BRLarmor},
the magnetic field in our case is parallel instead of orthogonal to the direction of motion.
We prove that our definition can reduce to the traversal time for a free particle and also give meaningful results
for a square barrier.
For example, the transmitted Larmor time through square barrier approaches the B$\ddot{\mathrm{u}}$ttiker-Landauer time in the opaque limit.
As our focus is the temporal behavior for a particle scattering off a linear barrier,
we degrade the proof in the appendix.
We hope with our extensive discussion on the various tunneling times and their relations,
linear barrier as another theoretical test ground (besides square barrier) in resolving the tunneling time issue
can be appreciated and may even draw the interest of experimental physicists.

The remainder of the paper is organized in the following way. In Sec.\ref{Prel}, we review the preliminary knowledge
in describing neutron's motion in the linear potential. At the end of this section, we also provide rough estimates on the time
scales involved in particular tunneling process based on the uncertainty principle and the semi-classical approximation.
In Sec.\ref{PhaseDW}, we discuss in detail about the calculations of phase and dwell times,
and the deviation of these times from the classical returning time (CRT).
In Sec.\ref{LarT}, we illustrate our definition of Larmor time and calculate its zero magnetic field limit for a linear barrier.
Interestingly, the limit is just the CRT. At last, we summarize our main results in Sec.\ref{Sumy}.

\section{Basic Theory}\label{Prel}
\subsection{A neutron in a linear potential}
Before discussing tunneling time, we briefly review the quantum description of neutron's motion
in the linear potential $V_b=m_{_G}gz$ \cite{BallQM}.
The general solution to the Schr$\ddot{\mathrm{o}}$dinger equation
\bea\label{NeutronSurf}
i\hbar\frac{\prt}{\prt t}\Phi(z,t)=\left[-\frac{\hbar^2}{2m_{_I}}\frac{\prt^2}{\prt z^2}+ m_{_G}gz\right]\Phi(z,t)
\eea is
$\Phi(z,t)=\int{dE}e^{-iEt/\hbar}\rho(E)\phi_E(z)$, where $\phi_E(z)$ is the solution of the
stationary Schr$\ddot{\mathrm{o}}$dinger equation
\bea\label{statSch}
\phi''_E(z)+\frac{2mE}{\hbar^2}(1-\frac{mg}{E}z)\phi_E(z)=0,
\eea
and $\rho(E)$ is the weighting factor for the Fourier expansion of the wave-packet $\Phi(z,t)$.
For simplicity, we assume the equality between the gravitational mass and inertial mass of neutron,
\ie, $m=m_{_G}=m_{_I}$.
From the dimensional constants $g$, $m$, a characteristic length scale
$L_c\equiv(2m^2g/\hbar^2)^{-1/3}$ ($L_c=5.866\mu$m for neutron) can be constructed.
Associatively, the characteristic momentum $p_c\equiv\hbar/L_c$ and energy
$E_c\equiv[(mg\hbar)^2/2m]^{1/3}$ can be defined.
With these dimensional constants, we can rewrite the equation (\ref{statSch}) with the dimensionless
variables $z_{_D}=z/L_c$ and $E_{_D}=E/E_c$. The general solution is
\bea
\phi_E(z_{_D})=c_1\mathrm{Ai}[z_{_D}-E_{_D}],
\eea
where we have abandoned the $\mathrm{Bi}[z-E_{_D}]$ branches as $\lim_{z_{_D}\rightarrow\infty}\phi_E(z_{_D})=0$.
To fix the undetermined constants $c_1$, a normalization condition is needed for bound state solutions and an additional
ansatz and continuity condition are needed for scattering state solutions.
According to different boundary conditions, the scattering and the bounded solutions
can be obtained respectively.
This paper is mainly to discuss the temporary behavior of the scattering state, so in the following, we
will shortly review the scattering state solution.
For details on the neutron gravitational bound states, see \cite{NGBS}\cite{ZXUCN}.
\begin{figure}
 \centering
 \hspace{0.2in}
 {\includegraphics[width=88mm]{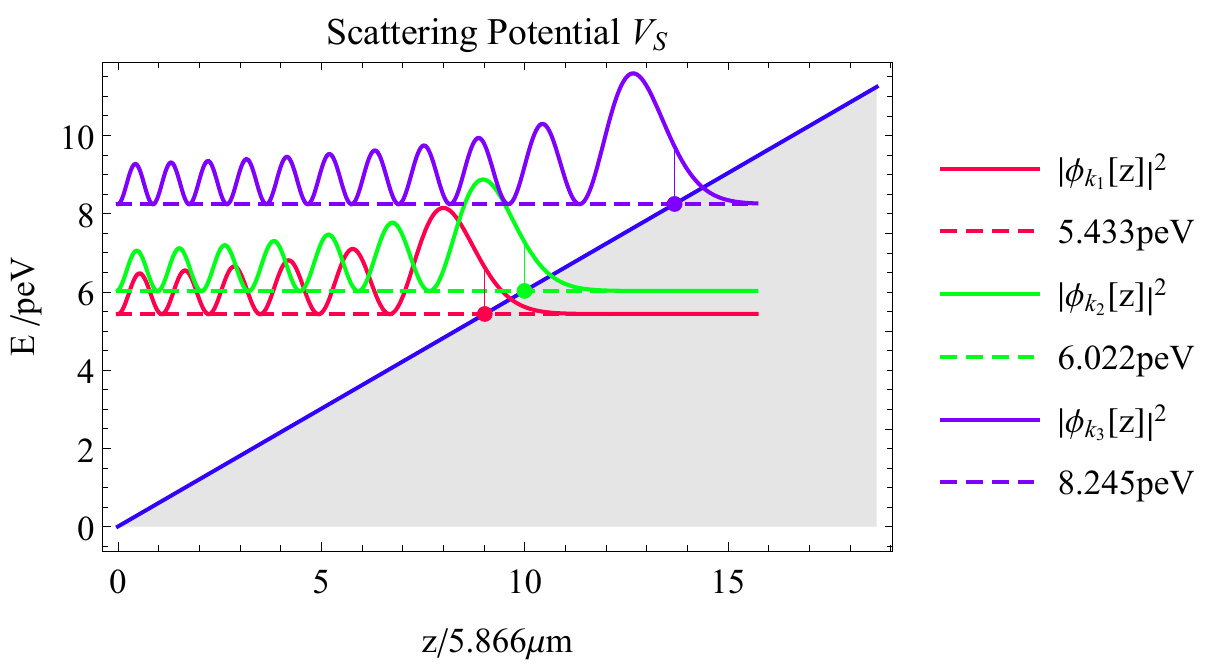}}
      \caption{Potential barrier $V_s[z]$ and the sample scattering wavefunctions.
    The red, green and purple curves correspond to the $|\phi_(k, z)|^2$
    with different $E_k=(\hbar k)^2/2m_I$, which are also shown in the corresponding colors with dashed horizontal lines.
    The three points are the cross points for each $E_k$ at the corresponding classical turning heights.}\label{scatter}
\end{figure}

\subsection{Scattering Solution}
We assume initially a beam of neutrons with a given energy is injected from the $z<0$ region,
and the effect of gravity can be ignored in the negative $z$-zone to simplify our discussion.
Consequently, the potential can be approximated as $V_s=mgz\Theta(z)$.
The general solution is still a linear superposition
\bea\label{WPacket}
\Phi(z,t)=\int_0^\infty{dk}\rho(k)e^{-iE_kt/\hbar}\phi(k,z),
\eea
where the energy $E_k=(\hbar k)^2/2m$, $\rho(k)$ is the weighting factor and $\phi(k, z)$ is the plane wave component
\bea\label{WaveZN}
\phi(k, z)={\Big\{}\begin{array}{c}
                          e^{ikz}+\mathcal{R}e^{-ikz},  ~~~~~~~~~~~~~z<0,\\
                          c_1\mathrm{Ai}[z/L_c-E_{_D}],~~~~~~~~~~ z>0.
                        \end{array}
\eea
From the continuity equation
\bea\label{ContinuCEqn}
1+\mathcal{R}=c_1\mathrm{Ai}[-E_{_D}],\quad
(1-\mathcal{R})ik=\frac{c_1}{L_c}\mathrm{Ai}'[-E_{_D}]
\eea
we get the coefficients
\bea\label{ScattCoe}&&
c_1=\frac{2ikL_c}{\mathrm{Ai}'[-E_{_D}]+ikL_c\mathrm{Ai}[-E_{_D}]},\\\label{RefCoe}&&
\mathcal{R}=\frac{\mathrm{Ai}[-E_{_D}]+\frac{i}{kL_c}\mathrm{Ai}'[-E_{_D}]}
{\mathrm{Ai}[-E_{_D}]-\frac{i}{kL_c}\mathrm{Ai}'[-E_{_D}]}.
\eea
From (\ref{RefCoe}), we can readily obtain $R=|\mathcal{R}|^2=1$,
since for any real argument $x$, $\mathrm{Ai}[x]\in\mathbb{R}$ and $\mathrm{Ai}'[x]\in\mathbb{R}$.
From the asymptotic form of $\mathrm{Airy}$ function at sufficiently (positive)
large $z$,
\bea\label{Asympt1}&&
\mathrm{Ai}[z]\sim\frac{e^{-\frac{2}{3}z^{\frac{3}{2}}}}{2\sqrt{\pi}z^{\frac{1}{4}}}
\sum_{n=0}^\infty\left[\frac{(-1)^n\Ga(n+\frac{5}{6})\Ga(n+\frac{1}{6})(\frac{3}{4})^n}{2\pi{n!}z^{\frac{3n}{2}}}\right]\nn&&
~~=\frac{e^{-\frac{2}{3}z^{\frac{3}{2}}}}{2\sqrt{\pi}z^{\frac{1}{4}}}\left[1-\frac{5}{48}z^{-\frac{3}{2}}
+\frac{385}{4608}z^{-3}+...\right],
\eea
we get the expected asymptotic decay behavior.
From the asymptotic behavior, we can define the barrier penetration length as $L_p\equiv{}L_c(1+k^2L_c^2)$,
which is the depth where the argument in the Airy function in Eq.~(\ref{WaveZN}), $z/L_c-E_{_D}=L_p/L_c-E_{_D}=1$.
In Fig.\ref{scatter}, we plot the relative probability density $|\phi(k, z)|^2$ with
the linear potential $V_S$.
From Fig.\ref{scatter}, we can readily see the small tails of $|\phi(k_i, z)|^2$ penetrating into the classical forbidden region (CFR),
and the amplitudes of $|\phi(k_i, z)|^2$ decrease at much lower $z$ than the classical turning heights $z_{_C}=E_k/mg$ for each $k_i$
($i=1,2,3$), indicated by the cross points of the dashed horizontal lines (representing the corresponding eigen-energies)
and the oblique solid blue line (representing the linear potential).
The small penetrating tails indicate that the incoming neutron has a small probability tunneling into the CFR.
Later we will see the advanced decreasing of wave amplitude at position lower than $z_{_C}$ and the penetration of wave amplitude into the CFR
conspire to match the total dwell time with the CRT.

\begin{table}[ht]
\begin{center}
\begin{tabular}{c|c|c|c|c|c|c|c|c}
\hline\hline
    Tunneling     &  ETI\cite{AttoHe}     &  NTLG     &  ATL\cite{AtomT} \\
 \hline
   $L$    &  0.591nm($\frac{I_0}{eF}$)  &  5.866$\mu$m($L_c$)  &  1.3$\mu$m(waist)  \\
   $m$             &  510.99keV($m_e$)   &  939.57MeV($m_n$)  &  81.69GeV($m_{^{87}\text{Rb}}$)  \\
   $E$              & 24.59eV   &  0.602peV($E_c$)  &  10.47peV(122nK)     \\
   $V$              & 24.59eV   & 6.624peV($mgL_p$)  &  15.51peV(180nK)   \\
\hline \hline
   $\tau_\mathrm{uc}$      &  48as   &  1.09ms  &  0.06ms    \\
   $\tau_\mathrm{sm}$      &  284as   &  0.55ms  &  0.39ms  \\
\hline \hline
   $\tau_\mathrm{exp}$      &  34as\cite{AttoHe}   &  ?  &  0.62ms\cite{AtomT}    \\
\hline\hline
\end{tabular}\caption{Simple estimations on time scales in various tunneling processes.
   In this table, $\tau_\mathrm{uc}$ is the estimate from uncertainty principle, $\tau_\mathrm{sm}$ is the semi-classical estimate,
   and $\tau_\mathrm{exp}$ comes from experimental measurements. Up to know, we don't know any measurement
   on neutron's tunneling delay in the surface gravity of the Earth.
   For briefness, ETI,  NTLG and ATL refer to electron tunneling ionization in helium, neutron tunneling in linear gravity,
   and atom tunneling through blue detuned laser field, respectively. }\label{timeE}
\end{center}
\end{table}
Before diving into any specific definition of tunneling time, we give simple estimates on the time scales involved in
three particular tunneling processes based on:
\bit
\item Uncertainty principle estimate, $\tau_\mathrm{uc}\sim\frac{\hbar}{\de E}$, where $\de E$ is the characteristic energy involved
      in a specific process;
\item Semi-classical estimate, $\tau_\mathrm{sm}\sim\frac{L}{\sqrt{2(V-E)/m}}$, where $L$ is the characteristic length,
      $m$ and $V-E$ are the mass and the negative kinetic energy of the tunneling particle.
\eit
Let's take the electron tunneling ionization in a hydrogen
atom as an example, where the ionization potential is $I_0=13.6$eV.
Given that the laser peak intensity is $2.9\times10^{14}\text{W}/\text{cm}^2$, the corresponding electric field strength is around $E=4.67\times10^{10}$V/m.
There are two characteristic lengths, the effective barrier width $d_1=I_0/(eE)=2.91{\AA}$ \cite{KELDYSH} and the Bohr radius $a_0=\hbar/(m_ec\al)\simeq0.53{\AA}$
(for the natural field strength $E=\al c\hbar/(ea_0^2)=5.14\times10^{11}$V/m, the effective barrier width $d_2\simeq{a_0}/2$).
Here $-e,~m_e$ are the charge and mass of the electron, and $\al$ is the fine structure constant.
The uncertainty principle estimate gives $\tau_\mathrm{uc}\sim48$as, and the semiclassical estimate gives $\tau_\mathrm{sm}=d_1/\sqrt{2I_0/m}\sim134$as (for $d_2$, $\tau_\mathrm{sm}\sim12$as, roughly the same order
as $\tau_\mathrm{uc}$).
The estimations give roughly $10\sim100$as for the duration of tunneling ionization,
comparable to the measurement performed on Helium \cite{AttoHe}.
For other tunneling processes, we summarize the simple estimates in Table \ref{timeE},
where in the last row, ``$?$" means that up to now, no corresponding result is known experimentally.

\section{Reflective phase time and dwell time}\label{PhaseDW}
\begin{figure}
 \centering
  {\includegraphics[width=75mm]{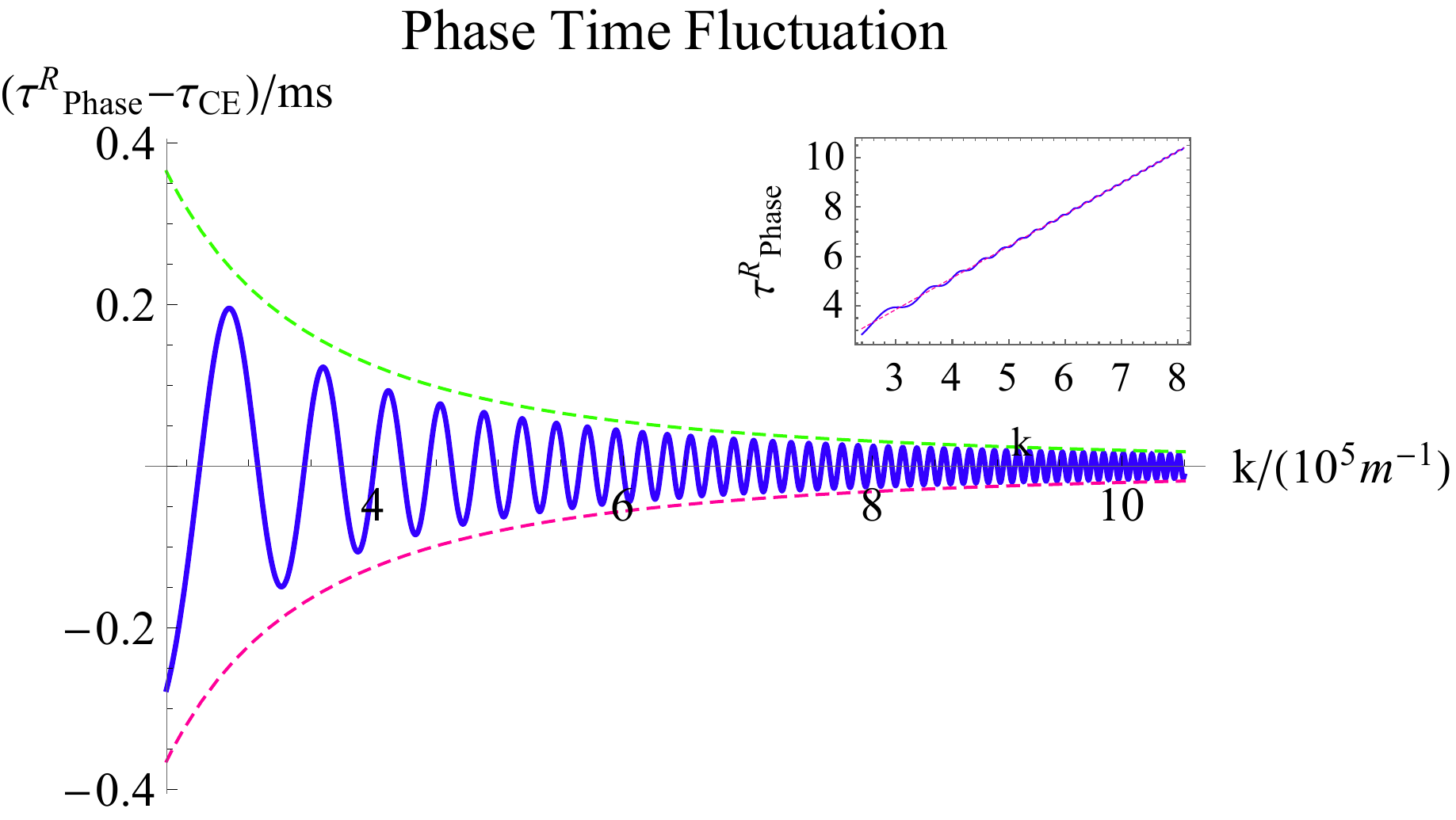}}
  \hspace{0.2in}
    \caption{~The quantum deviation of reflective phase time around the classical returning time,  $\tau^R_\mathrm{phase}-\tau_{_\mathrm{CE}}$, represented by the solid blue curve.
   The envelope fitting curves are given by $\pm\tau_\mathrm{coh}$, represented by the dashed red and green curves, respectively.
   The inset on the upper right corner shows the reflective phase time $\tau^R_\mathrm{phase}$ and the classical returning time $\tau_{_\mathrm{CE}}$, represented by solid blue curve and dashed red line, respectively. }\label{RCtau}
\end{figure}

The Wigner phase time (or Eisenbud-Wigner time, or group time delay) \cite{EPW}\cite{Smith}
is to follow the peak evolution of the wave packet and can be determined by the stationary phase method.
As there is no obvious causal connection between the peaks of incoming packet and the transmitted one \cite{BITLM},
and a simple barrier behaves like a high-energy components filter \cite{HLK2019}, phase time cannot directly correspond to the true time duration in tunneling \cite{Winful2006}.
However, it can still characterize the time scales in a quantum process \cite{ZXRT2015}, and can clearly
demonstrate the self-interference between the incoming and reflective partial waves. 
By definition, a stringent and satisfactory treatment should start with a wave packet with finite width, however,
for simplicity, we choose the monochromatic approximation, \ie, $\rho(k)\simeq\de(k-k_0)$ instead.
In the following, we stick to this assumption and drop the low index $0$ in $k_0$.
However, we have to keep in mind that, strictly speaking, all the finite tunneling delays obtained below
are specific averages over the peculiar weighting function $\rho(k)\simeq\de(k-k_0)$,
and this is sufficient for most illustrative purposes.
As there is no transmitted neutron for an infinite high barrier, indicated by $R=1$,
we only need to calculate the reflective phase time $\tau^R_\mathrm{phase}\equiv2\hbar\frac{\prt\theta_R}{\prt{E}}$.
According to Eq.(8), we can reform the reflective amplitude as $\mathcal{R}\equiv\exp[i2\theta_R]$, where $\theta_R\equiv\arctan\left[\mathrm{Ai}'[-E_{_D}]/(kL_c\mathrm{Ai}[-E_{_D}])\right]$ and $E_{_D}=k_{_D}^2\equiv(kL_c)^2$.
With a tedious calculation, we can get
\bea\label{TRphase}&&
\tau^R_\mathrm{phase}
=\frac{2m}{\hbar{}k}\frac{\prt\theta}{\prt{k}}\nn&&~~
=\frac{2L_c}{v(k)}\frac{2E_{_D}\left(\mathrm{Ai}'[y]^2-\mathrm{Ai}[y]\mathrm{Ai}''[y]\right)-\mathrm{Ai}'[y]\mathrm{Ai}[y]}
{\mathrm{Ai}'[y]^2+E_{_D}\mathrm{Ai}[y]^2},\nn
\eea
where $v(k)\equiv\hbar{k}/m$  and $y\equiv-E_{_D}$.
To clearly demonstrate the quantum induced time delay, we subtract the classical returning time $\tau_{_\mathrm{CE}}\equiv2v(k)/g$ from the reflective phase time $\tau^R_\mathrm{phase}$, and the difference $\tau^R_\mathrm{phase}-\tau_{_\mathrm{CE}}$ is plotted in Fig.\ref{RCtau}.
As an illustration, we also plot $\tau^R_\mathrm{phase}$ (solid blue curve) and $\tau_{_\mathrm{CE}}$ (dashed red line) with respect to wave number $k$ in the upper right inset.
From the inset, we see that for large $k$ (corresponds to energetic neutron),
$\tau^R_\mathrm{phase}$ matches with $\tau_{_\mathrm{CE}}$ very well,
while for small $k$ with the corresponding de Broglie wave length comparable to $L_c$,
the quantum deviation $\tau^R_\mathrm{phase}-\tau_{_\mathrm{CE}}$ becomes clear,
as evidently shown by the solid blue curve in Fig.\ref{RCtau}.
Clearly, with the increase of neutron wave number $k$, the oscillating amplitude of
$\tau^R_\mathrm{phase}-\tau_{_\mathrm{CE}}$ decreases quickly.

We can interpret the oscillating behavior as a manifestation of the self-interference
between incoming and reflective partial waves \cite{Fertig}.
The de Broglie wave length of a neutron with wave number $k$ is $\lambda_k=2\pi/k$.
Only two partial waves with comparable de Broglie wave lengths can interfere coherently,
so a rough estimate of the time scale is $\tau_\mathrm{coh}=\al\lambda_k/v(k)=\al\frac{2m\pi}{\hbar k^2}$,
where $\al$ is a $\mathcal{O}(1)$ free factor inserted for numerical fitting.
We choose $\al=1/5$ to fit the quantum deviation $\tau^R_\mathrm{phase}-\tau_{_\mathrm{CE}}$.
The rough estimate $\tau_\mathrm{coh}$ fits the envelope of the oscillating phase time delay very well, see the dashed red and green fitting curves in Fig.\ref{RCtau}. Therefore the excellent fitting supports our self-interference interpretation \cite{Fertig}.

\begin{figure}
 \centering
  \subfigure[Various contributions to dwell time]{\includegraphics[width=78mm]{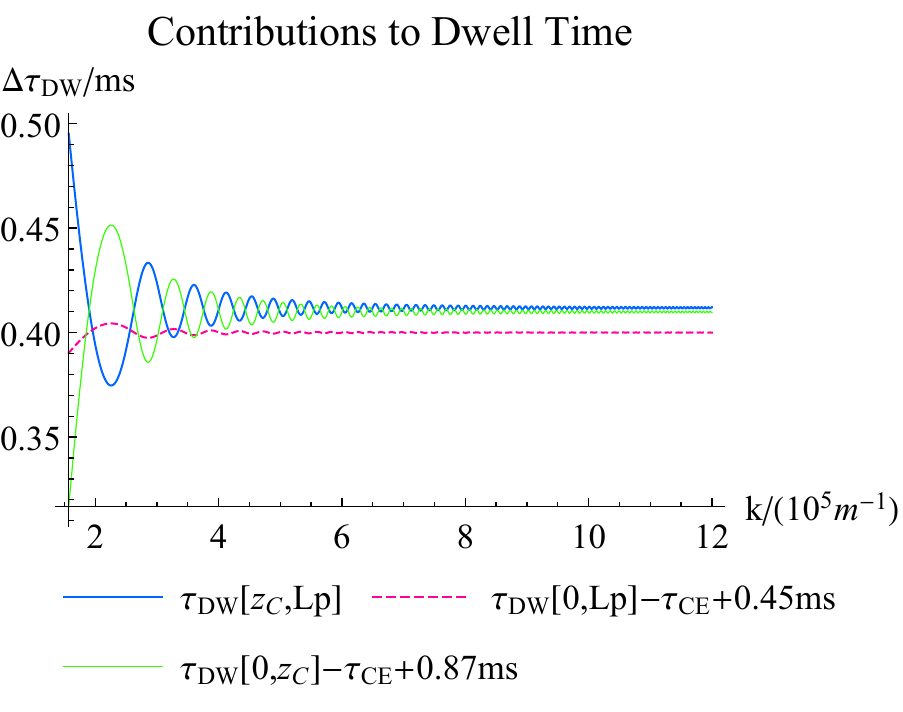}\label{DWFluC}}
  \subfigure[~ Self-interference time $-\tau_{_{\mathrm{IF}}}$ ]{\includegraphics[width=75mm]{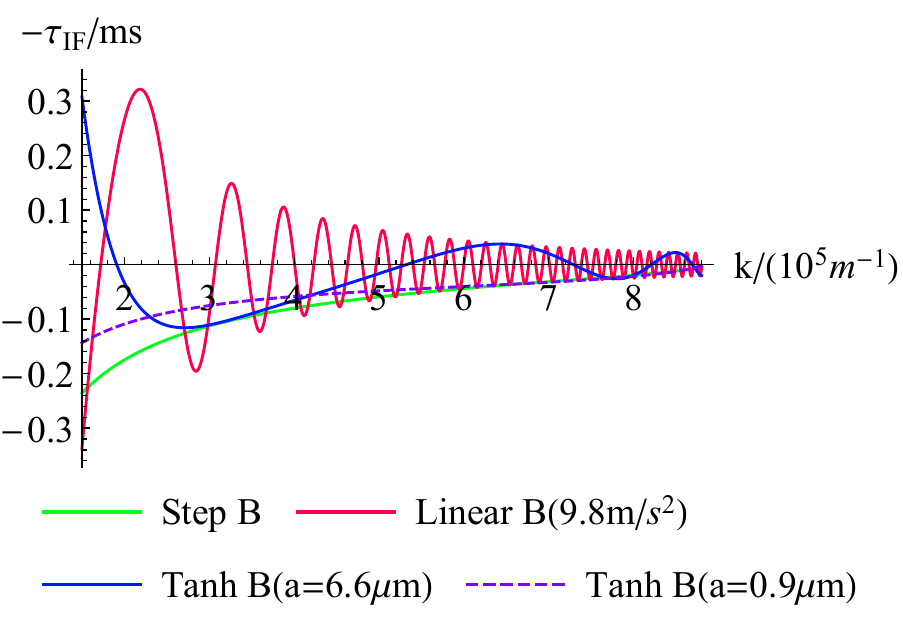}\label{tauIF}}
     \caption{ a. Deviations of dwell time in various region to $\tau_{_{\mathrm{CE}}}$ and the dwell time in the classical forbidden region (CFR) $\tau_{_{\mathrm{DW}}}[z_c,L_p]$. The $\tau_{_{\mathrm{DW}}}[z_c,L_p]$ is represented by the solid blue curve. The deviations $\tau_{_\mathrm{DW}}[0,L_P]-\tau_{_{\mathrm{CE}}}$ and $\tau_{_\mathrm{DW}}[0,z_C]-\tau_{_{\mathrm{CE}}}$ are plotted with dashed red and solid green curves, respectively. For comparison convenience, they are constantly shifted by $0.45$ms and +$0.87\mathrm{ms}$, respectively. b. Self-interference time delay for various semi-infinite barriers. The $-\tau_{_{\mathrm{IF}}}$ for gravitational linear barrier is shown
     in solid red curve, while for step barrier, $-\tau_{_{\mathrm{IF}}}$ is shown in solid green curve. For the barrier $V=\frac{A}{2}[1+\tanh(x/a)]$, the time for $a=0.9\mu$m and $a=6.6\mu$m are represented by the dashed purple and solid blue curves, respectively.
     }
\end{figure}

Next we turn to the dwell time, which was first introduced in Ref. \cite{Smith}. Unlike phase time,
dwell time is a positively defined quantity, averaged over all scattering channels.
The indistinguishability between reflected and transmitted channels means dwell time is better viewed as
a lifetime or a storage time rather than a traversal time.
However, in the case of a linear barrier, as all particles including the ones penetrated into
the barrier are finally reflected, indicated by $|\mathcal{R}|=1$, dwell time does encode the tunneling time delay.
According to Ref. \cite{Buttiker1983}, dwell time can be defined as
\bea\label{dwellT}
\tau_{_{\mathrm{DW}}}[-z_L,z_R]\equiv\frac{m}{\hbar k}\int_{-z_L}^{z_R}dz|\phi(k,z)|^2,
\eea
where the positive $z_L,~z_R$ can be chosen as the characteristic length $L_c$ and the penetration depth $L_p$, respectively.
The integral in (\ref{dwellT}) is given by
{\small
\bea&&
\int_{-L_c}^{L_p}dz|\phi(k,z)|^2=\int_{-L_c}^0dz\left[2+(\mathcal{R}e^{-2ikz}+c.c)\right]\nn&&~~~~~~
+|c_1|^2\int_{0}^{L_p}dz\mathrm{Ai}^2[z/L_c-E_{_D}]=
\nn&&~~~~~~
L_\mathrm{eff}+\frac{4}{\mathrm{Ai}^2+(\frac{\mathrm{Ai}'}{kL_c})^2}\int_{0}^{L_p}dz\mathrm{Ai}^2[z/L_c-E_{_D}],
\eea}
where $\mathrm{Ai},~\mathrm{Ai}'$ represent $\mathrm{Ai}[-E_{_D}],~\mathrm{Ai}'[-E_{_D}]$ respectively,
\bea\label{Leff}&&
L_\mathrm{eff}\equiv
\frac{[\mathrm{Ai}^2-(\frac{\mathrm{Ai}'}{kL_c})^2]\frac{\sin(2kL_c)}{k}+\frac{2\mathrm{Ai}'\mathrm{Ai}}{k^2L_c}[\cos(2kL_c)-1]}
{[\mathrm{Ai}^2+(\frac{\mathrm{Ai}'}{kL_c})^2]}\nn&&~~~~
+2L_c,
\eea
and $\int_{0}^{L_p}dz\mathrm{Ai}^2[z/L_c-E_{_D}]$ is given in (\ref{AiSI}).
Substituting all these terms into (\ref{dwellT}), we can get
{\small
\bea\label{dwellT1}&&
\tau_{_{\mathrm{DW}}}[-L_c,L_p]=\tau_{_\mathrm{CE}}\left\{1+\frac{\mathrm{Ai}^2[1]-\mathrm{Ai}'^2[1]}
{E_{_D}\mathrm{Ai}^2+(\mathrm{Ai}')^2}\right\}
+\frac{m_{_I}L_\mathrm{eff}}{\hbar k},\nn
\eea
}
where $\tau_{_\mathrm{CE}}=\tau_{_{\mathrm{DW}}}[0,+\infty]$ 
is the dwell time in the region of linear barrier.
\begin{figure}
 \centering
 \subfigure[~Dwell time $\tau_{_{\mathrm{DW}}}$ vs classical returning time $\tau_{_\mathrm{CE}}$]{\includegraphics[width=72mm]{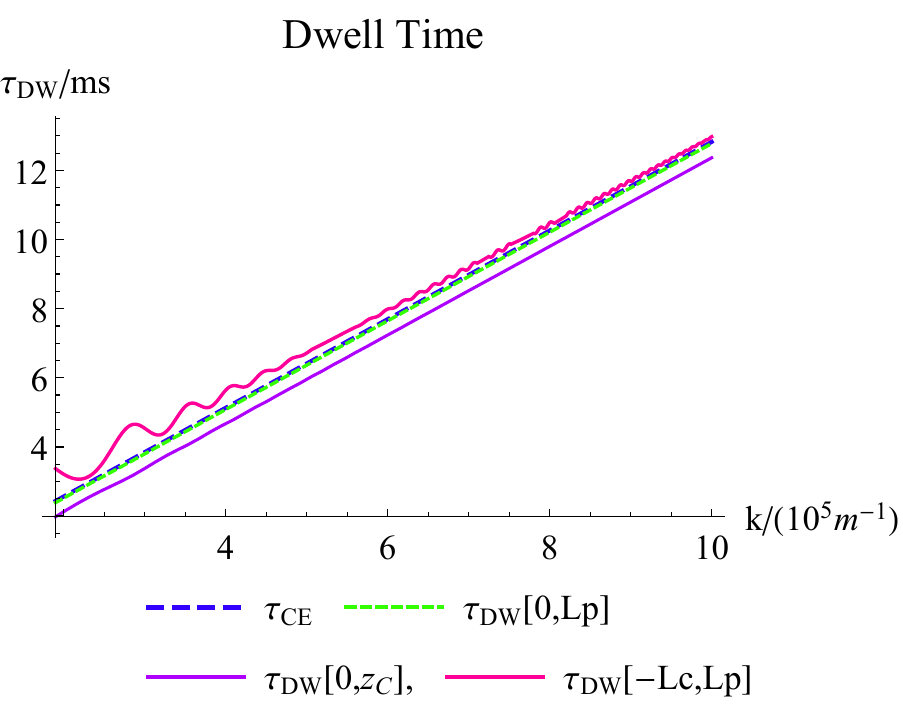}\label{DWtau1}}
  \subfigure[~ Dwell time fluctuation $\de\tau_{_{\mathrm{DW}}}$ ]{\includegraphics[width=75mm]{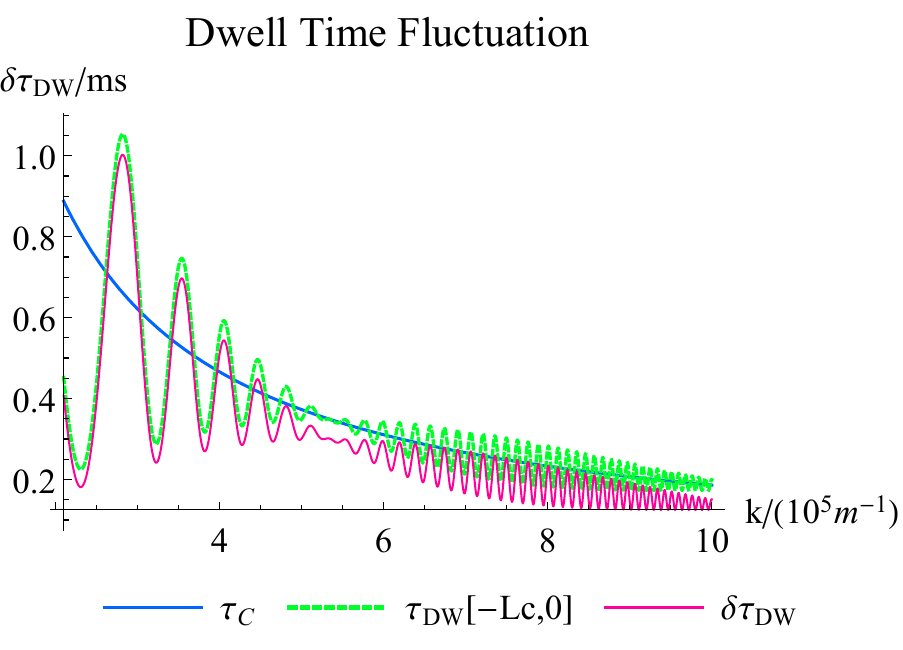}\label{tauDWFluc}}
     \caption{~a. Dwell time with respect to CRT, represented by the solid red and dashed blue curves respectively.
     The first part of $\tau_{_{\mathrm{DW}}}$ in (\ref{dwellT1}), $\tau_{_{\mathrm{DW}}}[0,L_p]$, and the dwell time in
     the classical expected region, $\tau_{_{\mathrm{DW}}}[0,z_C]$, are represented by the dashed green and solid purple curves, respectively.
     b. The second part of $\tau_{_{\mathrm{DW}}}$, $mL_\mathrm{eff}/(\hbar k)$, and the quantum deviation (or fluctuation) of $\tau_{_{\mathrm{DW}}}$ from the CRT, $\delta\tau_{_{\mathrm{DW}}}\equiv\tau_{_{\mathrm{DW}}}-\tau_{_{CE}}$, are represented by the dashed green and solid red curves, respectively. The solid blue curve represents $\tau_{_C}\equiv2mL_c/(\hbar k)$.
     }\label{DWtau}
\end{figure}
We plot the dwell time with respect to the classical returning time $\tau_{_\mathrm{CE}}$ in Fig.\ref{DWtau1},
where the dashed blue line and the solid red curve correspond to $\tau_{_\mathrm{CE}}$ and the dwell time in (\ref{dwellT1}), respectively.
To analyze the contributions of the first and second parts in (\ref{dwellT1}) to $\tau_{_{\mathrm{DW}}}[-L_c,L_p]$ (briefly referred as
$\tau_{_{\mathrm{DW}}}$ in the following),
we plot them separately in Fig.\ref{DWtau1} and Fig.\ref{tauDWFluc}, represented by the corresponding dashed green curves.
For later convenience, we also plot the dwell time in the classical expected region $[0,z_{_C}]$
in Fig\ref{DWtau1}, see the solid purple curve.
Similar to the phase time shown in the inset of Fig.\ref{RCtau}, we see that the dwell time $\tau_{_{\mathrm{DW}}}$ also gets closer to $\tau_{_\mathrm{CE}}$ with increasing $k$, see Fig.\ref{DWtau1}.
This confirms our previous observation that the deviation of the behavior for a vertically injected neutron in the Earth gravitational
field from the classical prediction tends to $0$ with increasing kinetic energy.
From the nearly overlap of the dashed green line, $\tau_{_\mathrm{DW}}[0,L_p]$, with the dashed blue line, $\tau_{_\mathrm{CE}}$, in Fig.\ref{DWtau1}, we see that the dominant part of $\tau_{_{\mathrm{DW}}}$ is $\tau_{_\mathrm{DW}}[0,L_p]$,
while the quantum fluctuation is largely due to $\tau_{_\mathrm{DW}}[-L_c,0]\equiv mL_\mathrm{eff}/(\hbar k)$, the dwell time in the region
where the gravitational interaction is ignored in our simple approximation.
In comparison, we plot the quantum deviation $\de\tau_{_\mathrm{DW}}\equiv\tau_{_{\mathrm{DW}}}-\tau_{_{CE}}$ (the solid red curve) with
$\tau_{_\mathrm{DW}}[-L_c,0]$ (the dashed green curve) in Fig.\ref{tauDWFluc},
where the two curves nearly coincide, and the tiny discrepancy is due to $\tau_{_\mathrm{DW}}[0,L_p]-\tau_{_{CE}}$, represented by
the dotted red curve (slightly shifted by $0.45$ms) in Fig.\ref{DWFluC}.
The solid blue curve in Fig.\ref{tauDWFluc} represents $\tau_{_C}\equiv2mL_c/(\hbar k)$, and
the factor 2 is to count for the time of return of the incoming neutron.
Clearly, $\tau_{_\mathrm{DW}}[-L_c,0]$ oscillates around $\tau_{_C}$,
a naive classical estimate of the returning delay for a free particle.
To show the important contribution of the classical forbidden region (CFR) $[z_c,L_p]$ to $\tau_{_\mathrm{DW}}$,
we plot $\tau_{_\mathrm{DW}}[z_c,L_p]$ in Fig.\ref{DWFluC}, see the solid blue curve.
To facilitate the comparison, we also plot the constantly shifted deviations of $\tau_{_\mathrm{DW}}[0,z_{_C}]-\tau_{_{CE}}$ and $\tau_{_\mathrm{DW}}[0,L_p]-\tau_{_{CE}}$ in Fig.\ref{DWFluC}, represented by the dashed green and dashed red
curves, respectively.
The shifted time constants are shown in the legends below the figure.
By comparing the oscillation amplitudes of the dashed green and dashed red curves in Fig.\ref{DWFluC}
(the constant time shifts are not very relevant here),
we find that $\tau_{_\mathrm{DW}}[0,L_p]$ is much closer to $\tau_{_{CE}}$ than $\tau_{_\mathrm{DW}}[0,z_{_C}]$,
which can also be seen transparently in Fig.\ref{DWtau1},
where there is a nearly constant gap between the dash green line and the solid purple line, representing $\tau_{_\mathrm{DW}}[0,L_p]$
and $\tau_{_\mathrm{DW}}[0,z_{_C}]$, respectively.
So without the contribution from the CFR, represented by the solid blue curve in Fig.\ref{DWFluC},
the intuitively more ``classical" returning time $\tau_{_\mathrm{DW}}[0,z_{_C}]$ deviates from $\tau_{_\mathrm{CE}}$ by an
indispensable discrepancy, which is a constant $-4mL_c^2\pi/(\hbar3^{2/3}\Gamma[1/3]^2)$ when $k\rightarrow+\infty$
[can also be seen from the rapidly decreasing oscillating amplitude of $\tau_{_\mathrm{DW}}[z_{_C},L_p]$,
represented by the solid blue curve in Fig.\ref{DWFluC}].

As elegantly displayed by the manipulation of the stationary Schrodinger equation in Ref.\cite{Winful0304},
there is a relation between dwell time and phase time.
The direct application of the relation in \cite{Winful0304} does not work, as there is no asymptotically transmitted region in the far right side for a linear barrier. However, with a small alteration, the relation becomes
\bea\label{RelaTPDT}
\tau_{_\mathrm{DW}}[0,+\infty]=\frac{m}{\hbar\,k^2}\mathrm{Im}[\mathcal{R}]+\tau^R_\mathrm{phase}.
\eea
The formal simplicity of this relation strongly depends on the specifically chosen integration interval $[0,+\infty]$,
where the wave function and its derivatives simply vanishes at $+\infty$
(Actually, even replacing $+\infty$ in $\tau_{_\mathrm{DW}}[0,+\infty]$ by a finite large coordinate $z_R>0$
in the left hand side of (\ref{RelaTPDT}), it still holds true to an good approximation,
due to the exponential decay of wave function in the barrier region).
Other choice of integration interval may generate a relation looking more complicated.
For example, if choosing $z_R=L_p$ instead of $+\infty$ or any other large value, there will be an additional term
\bea\label{AddiDW}&&
\tau_{_{L_p}}=\frac{m}{\hbar\,k}\frac{4L_cE_{_D}\left(\mathrm{Ai}^2[1]-\mathrm{Ai}'^2[1]\right)}
{\mathrm{Ai}'^2[-E_{_D}]+E_{_D}\mathrm{Ai}^2[-E_{_D}]},
\eea
which is just $\tau_{_\mathrm{DW}}[0,L_p]-\tau_{_\mathrm{CE}}$, and is represented by the dotted red curve
in Fig.\ref{DWFluC}.
Eq. (\ref{RelaTPDT}) is also applicable to barriers with $|\mathrm{R}|=1$,
and the term $\tau_\mathrm{IF}\equiv-\frac{m}{\hbar\,k^2}\mathrm{Im}[\mathcal{R}]$
is called self-interference delay \cite{Winful2006}, originated from the overlap between incoming and reflective partial waves,
and is very sensitive to barrier shape.
As an illustration, we plot the corresponding $-\tau_{_\mathrm{IF}}$ in Fig. \ref{tauIF} for three different semi-infinite barriers:
the step barrier $V=A\,\theta(x)$, the $\tanh$-like barrer $V=\frac{A}{2}[1+\tanh(x/a)]$ \cite{ZXHH} and the linear barrier $V=mgx\theta(x)$.
To facilitate the comparison, we set $A=27mgL_c$ for step barrier and $\tanh$-like barrers \cite{ZXHH},
and we choose two different length parameters, $a=6.6\mu$m and $a=0.9\mu$m for the latter.
In comparison, the barrier with $a=0.9\mu$m is much steeper than $a=6.6\mu$m and resembles more to the step barrier.

From Fig.\ref{tauIF}, we see that the self-interference effect is more significant for low energy
particles, and becomes more evident for smooth and gentle barriers (linear barrier and $a=6.6\mu$m $\tanh$-like barrier,
see the solid red and solid blue curves, respectively) than for steep barriers
(step barrier and $a=0.9\mu$m $\tanh$-like barrier, see the solid green and dashed purple curves, respectively),
as expected.
Interestingly, as $\tau_{_{\mathrm{DW}}}[0,+\infty]=\tau_{_\mathrm{CE}}=4mkL_c^3/\hbar$ for linear barrier,
$\tau_{_\mathrm{IF}}$ is exactly the phase time delay shown in Fig.\ref{RCtau},
and confirms our assertion that the oscillating phase time delay is due to the self-interference.
Since $\tau_{_\mathrm{IF}}$ is sensitive to barrier shape, it also provides a clue of why dwell time is
barrier shape sensitive \cite{TTWM2016} from the relation (\ref{RelaTPDT}).

At last, we also note that the time estimate from the WKB approximation due to the CFR, $\tau_{_\mathrm{WKB}}\equiv\int_{z_c}^{L_p}dz\frac{2}{g}\sqrt{2(mgz-E)}=2\sqrt{2L_c/g}\simeq2.19$ms,
is roughly the same order as $\de\tau_{_\mathrm{DW}}$ for small $k$,
and is consistent with the rough estimate in Table \ref{timeE}.

\section{Larmor time and Neutron in an external magnetic field}\label{LarT}
\begin{figure}
\centering
  \includegraphics[width=75mm]{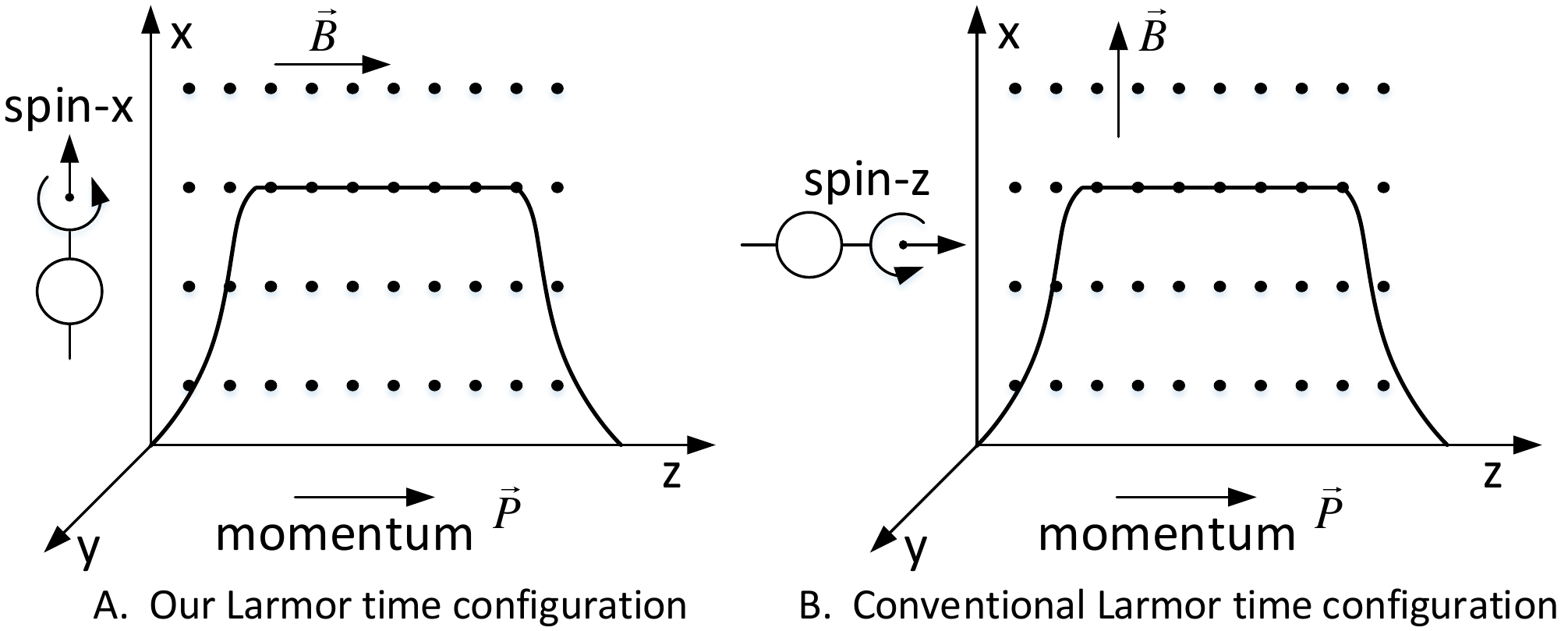}
      \caption{The sketch map of the spin and magnetic field configurations for conventional definition and our definition of Larmor times,
      shown in right and left, respectively. The dotted region indicates the region with barrier is covered by weak magnetic field $\vec{B}$.
      }\label{SMapLM}
\end{figure}

In analogy with the attoclock, where a highly circularly polarized electric field rotating on
the plane orthogonal to its direction of motion plays the role of a hand on the face of a clock,
the neutron spin can also act as a clock pointer.
The picture is that, as neutron carries non-zero magnetic moment, its spin precesses in an external magnetic field.
This internal degree of freedom acts as a ``pointer"
and the precession angle with respect to the initial spin measures the time elapsed during which the neutron
is in the region covered by magnetic field.
The spin precession is well-known as the Larmor precession, and hence the measured time is called the Larmor time \cite{BRLarmor}.
Larmor time has been discussed extensively in the literature \cite{BRLarmor}\cite{Yamada2004}\cite{TTWM2016}\cite{Buttiker1983},
and has an intimate connection with the complex time obtained by the path integral approach \cite{Sokolovski}.
Different from the conventional definition of Larmor time, in our approach the external magnetic field is
defined to be parallel instead of orthogonal to the direction of motion,
whereas the initial spin polarization is still orthogonal to the magnetic field to allow spin precession.
To illustrate the distinction, the initial spin, momentum and magnetic field for both definitions
are shown in Fig. \ref{SMapLM}.
We find that for a free particle, our definition reduces to the classical traversal time $L/v$,
where $L$ is the length of the space interval and $v$ is the speed of incoming particle.
For a rectangular barrier, the definition also gives a meaningful transmitted and reflected Larmor time.
As our goal is to discuss the tunneling time of a linear barrier, we degrade the Larmor time for the motion
through the free space and a rectangular barrier to the Appendix \ref{FSLarmorT}.

For a linear potential, to implement a Larmor clock, we introduce a magnetic field in the region $z\in[0,z_u]$, 
where $z_u\gg z_c=E/mg$. The Hamiltonian is given by
{
\bea&&
\hat{H}_{_{EG}}=\frac{\hat{\vec{p}}^2}{2m}+mgz\Theta(z)
-\vec{\mu}\cdot\vec{B}\Theta(z)\Theta(z_u-z),
\eea
}
where $\vec{\mu}\equiv\hbar\frac{\mu_N\tilde{g}}{2}\vec{\sigma}$ is the magnetic moment,
$\tilde{g}=-3.826085$ is the neutron Land\'e g-factor and $\mu_N$ is the nuclear magneton.
For simplicity, suppose the magnetic field is homogeneous such that $\vec{B}=B_0\vec{n}$,
and the non-relativistic neutron wave-function is described by a two-component
Pauli-spinor $\phi\equiv(\chi,~\eta)^T$.
Defining $\tilde{\mu}_N\equiv\mu_{_N}m_{_I},~b_0\equiv{B_0L_C^2/\hbar}$,
we can recast the static eigen-equation in the region $0<z<z_u$
into the dimensionless form
\bea
\tilde{\phi}''(z_{_D})+\left[E_{_D}(1-\frac{z_{_D}}{E_{_D}})+\tilde{g}\tilde{\mu}_Nb_0\vec{\si}\cdot\vec{n}\right]\tilde{\phi}(z_{_D})=0.
\eea
Assume the upward going neutron is initially polarized in a horizontal direction, \eg, being initially prepared in the eigen-state of $\sigma_x$, and the magnetic field is along the direction $\vec{n}\equiv(\sin\theta\cos\phi,\sin\theta\sin\phi,\cos\theta)$.
We keep a general $(\theta,\,\phi)$ temporarily, and at later stage we will set $\theta=0$ for results checking convenience.
Imposing the continuity condition at $z=0$ with the ansatz
\bea\label{ansatz1}
\phi(z)=\left\{\begin{array}{c}
          \frac{1}{\sqrt{2}}\left[\left(
                     \begin{array}{c}
                       e^{-i\alpha} \\
                       1 \\
                     \end{array}
                   \right)e^{ikz}+ \left(
                     \begin{array}{c}
                       \mathcal{R}_+ \\
                       \mathcal{R}_- \\
                     \end{array}
                   \right)e^{-ikz}\right],~~~z<0,
          \\
          U\left(
             \begin{array}{c}
               c_+\mathrm{Ai}[\frac{z}{L_c}-E_{_D}-\tilde{g}\tilde{\mu}_Nb_0] \\
               c_-\mathrm{Ai}[\frac{z}{L_c}-E_{_D}+\tilde{g}\tilde{\mu}_Nb_0] \\
             \end{array}
           \right), ~~~~z\geq0,
        \end{array}\right.
\eea
where
$U\equiv\left(
          \begin{array}{cc}
            \cos(\theta/2) & \sin(\theta/2)e^{-i\phi} \\
            \sin(\theta/2)e^{i\phi} & -\cos(\theta/2) \\
          \end{array}
        \right)$,
we get the coefficients below
\bea\label{Coeff}&&
c_+=\frac{\sqrt{2}[e^{-i\alpha}\cos\frac{\theta}{2}+e^{-i\phi}\sin\frac{\theta}{2}]}
{\mathrm{Ai}[-E_{_D}(1+r)]-\frac{i}{k_{_D}}\mathrm{Ai}'[-E_{_D}(1+r)]},\\&&
c_-=\frac{-\sqrt{2}[\cos\frac{\theta}{2}-e^{i(\phi-\alpha)}\sin\frac{\theta}{2}]}
{\mathrm{Ai}[-E_{_D}(1-r)]-\frac{i}{k_{_D}}\mathrm{Ai}'[-E_{_D}(1-r)]},
\eea
\bea\label{RCoeff}&&
\mathcal{R}_+=\left\{\frac{i}{k_{_D}}(\mathrm{Ai}'_+\mathrm{Ai}_--\mathrm{Ai}'_-\mathrm{Ai}_+)
[e^{-i\alpha}\cos\theta+e^{-i\phi}\sin\theta]\right.\nn&&~~
\left.+e^{-i\alpha}(\mathrm{Ai}_+\mathrm{Ai}_-+\frac{1}{k_{_D}^2}\mathrm{Ai}'_+\mathrm{Ai}'_-)\right\}/
\left[(\mathrm{Ai}_+-\frac{i}{k_{_D}}\mathrm{Ai}'_+)\right.\nn&&~~~~
\left.\cdot(\mathrm{Ai}_--\frac{i}{k_{_D}}\mathrm{Ai}'_-)\right],\\&&
\mathcal{R}_-=\left\{\frac{i}{k_{_D}}(\mathrm{Ai}'_-\mathrm{Ai}_+-\mathrm{Ai}'_+\mathrm{Ai}_-)
[\cos\theta-e^{i(\phi-\alpha)}\sin\theta]\right.\nn&&~~
\left.+(\mathrm{Ai}_+\mathrm{Ai}_-+\frac{1}{k_{_D}^2}\mathrm{Ai}'_+\mathrm{Ai}'_-)\right\}/
\left[(\mathrm{Ai}_+-\frac{i}{k_{_D}}\mathrm{Ai}'_+)\right.\nn&&~~~~
\left.\cdot(\mathrm{Ai}_--\frac{i}{k_{_D}}\mathrm{Ai}'_-)\right],
\eea
where we have defined $r\equiv\frac{\tilde{g}\tilde{\mu}_Nb_0}{E_{_D}}=(\tilde{g}\mu_{_N}m_{_I}B_0)/(k^2\hbar)$,
$\mathrm{Ai}_{\pm}\equiv\mathrm{Ai}[-E_{_D}(1\pm r)]$ and $\mathrm{Ai}'_{\pm}\equiv\mathrm{Ai}'[-E_{_D}(1\pm r)]$.

At this stage, we set $\theta=0$.
Since the incoming neutron is in spin state $\frac{1}{\sqrt{2}}(e^{-i\alpha},1)^T$,
by comparing the spin state of the reflected neutron,
we can read out the elapsed time through the spin precession angle
{\small
\bea\label{RotAng}
\Theta_R=2\left\{\tan^{-1}\left[\frac{(\mathrm{Ai}'_+\mathrm{Ai}_--\mathrm{Ai}'_-\mathrm{Ai}_+)}
{(k_{_D}\mathrm{Ai}_+\mathrm{Ai}_-+\frac{1}{k_{_D}}\mathrm{Ai}'_+\mathrm{Ai}'_-)}\right]\mathrm{mod}~\pi\right\},
\eea}
for details, see Appendix \ref{Methods}. Note the modulus of $\pi$ in (\ref{RotAng}) is only for mathematical rigor
and not necessary when the magnetic field strength $B_0$ is sufficiently small.
The magnetic field $B_0$-dependent Larmor time can be defined as
{\small
\bea\label{LarmorT0}&&
\tau_{\mathrm{Lar}}\equiv\frac{\Theta_R}{\omega_{\mathrm{Lar}}}=\frac{2m}{k^2r\hbar}\tan^{-1}\left[-\frac{G}{F}\right]
=\frac{-2}{\ga_{_N}B_0}\left[\frac{G}{F}\right],
\eea
}
where $F\equiv[\mathrm{Ai'}_+\mathrm{Ai'}_-+k_{_D}^2\mathrm{Ai}_-\mathrm{Ai}_+]$, $G\equiv[\mathrm{Ai}_+\mathrm{Ai'}_--\mathrm{Ai}_-\mathrm{Ai'}_+]k_{_D}$,
$\ga_{_N}=\tilde{g}\mu_N=1.832\times10^8\mathrm{rad}/(\mathrm{s}\mathrm{T})$ is the gyromagnetic ratio
of neutron and $\omega_{\mathrm{Lar}}\equiv\ga_{_N}B_0=r\hbar k^2/m$ is the Larmor frequency.
Clearly, our definition can be also applied to other potentials.
For the case of a finite barrier such as a rectangular barrier or a free region, it can have transmitted Larmor time as well.
In that case, $\Theta_R$ in (\ref{LarmorT0}) has to be replaced by the spin precession angle of the corresponding
transmitted partial wave with respect to the incoming partial wave, see Appendix \ref{FSLarmorT}.

\begin{figure}
\centering
  \includegraphics[width=70mm]{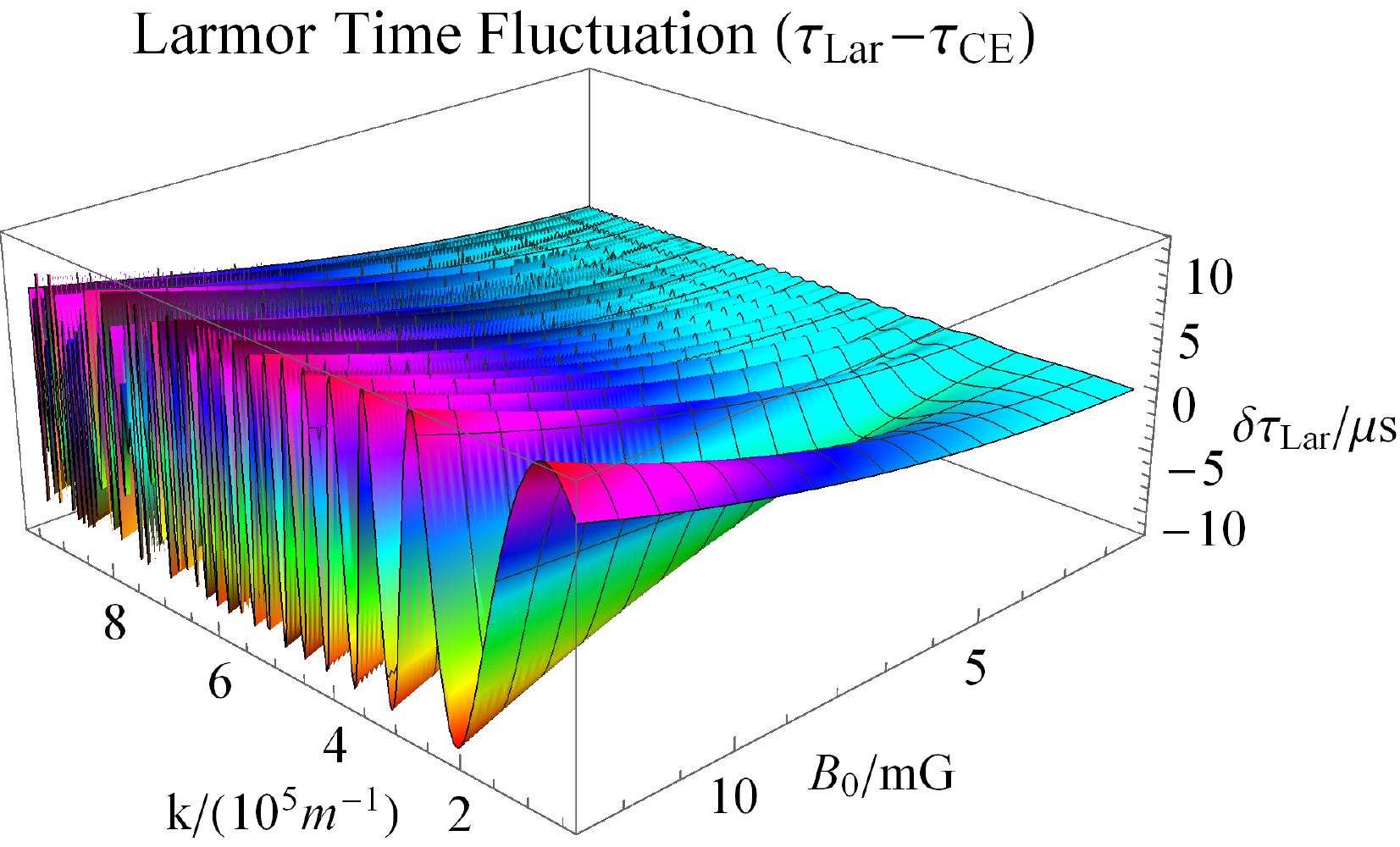}
      \caption{Deviation of Reflective Larmor time. The two-dimensional surface displays the deviation of $\delta\tau_{_{Lar}}\equiv\tau_{_\mathrm{Lar}}-\tau_{_\mathrm{CE}}$ with respect to wave number $k$ and the magnitude of
      weak magnetic field $B_0$. The magnetic field is in unit of $1\mathrm{mG}=10^{-7}$T, and the time deviation is in unit of $\mu$s. }\label{QFLTr}
\end{figure}

Interestingly, we find in the $B_0\rightarrow0$ limit (hence $r\rightarrow0$),
\bea\label{LarmorT}&&
\lim_{r\rightarrow0}\tau_{_\mathrm{Lar}}=\frac{2m}{k^2\hbar}
\frac{d}{dr}\left[\tan^{-1}(-G/F)\right]\nn&&~~
=\lim_{r\rightarrow0}\frac{2m\,k_{_D}^3}{k^2\hbar}\left\{2+r\frac{G\tilde{G}}{[F^2+G^2]}\right\}\nn&&~~
=\frac{4m\,kL_c^3}{\hbar}=\tau_{_\mathrm{CE}},
\eea
where $\tilde{G}\equiv[\mathrm{Ai}_+\mathrm{Ai'}_-+\mathrm{Ai}_-\mathrm{Ai'}_+]k_{_D}$.
This is clearly demonstrated in Fig.\ref{QFLTr}, where we plot the deviation of $\tau_{_\mathrm{Lar}}$ from the $\tau_{_\mathrm{CE}}$, $\delta\tau_{_\mathrm{Lar}}\equiv\tau_{_\mathrm{Lar}}-\tau_{_\mathrm{CE}}$.
In Fig.\ref{QFLTr}, we see that as the weak magnetic field decreases toward zero, $\tau_{_\mathrm{Lar}}$ approaches
the $\tau_{_\mathrm{CE}}$ as close as possible, and the deviation oscillates more frequently with the increase of the wave number $k$.
Compared with those deviations of the phase and the dwell times from CRT,
the oscillating amplitude decreases more gently with increasing $k$.
Note that $\lim_{r\rightarrow0}\tau_{_\mathrm{Lar}}=\tau_{_\mathrm{CE}}$ is not a common feature for general potentials,
such as a rectangular barrier, where the reflective Larmor time does not reduce to the CRT in the zero magnetic field limit.
This can be viewed as another support that a linear barrier is virtue of study for shape sensitive tunneling times.
Further, this is the second case where a quantum sojourning time agrees well with the classical returning time $\tau_{_\mathrm{CE}}=2\hbar\,k/(mg)$ for a linear barrier.

Similar to conventional definition of Larmor times, we can also define
\bea\label{CLarYZ}
\tau^{sy}_{~L}\equiv\langle\,S_y\rangle_s/(\hf\hbar\omega_{_\mathrm{Lar}}),~~~ \tau^{sz}_{~L}\equiv\langle\,S_z\rangle_s/(\hf\hbar\omega_{_\mathrm{Lar}})
\eea
associated with spin precession for an incoming spin-$\hf$ particle initially polarized in the $x$-direction.
The superscript or subscript $s=R,~T$, denote the reflective and transmitted scattering channels, respectively, for a general potential.
However, by intentionally aligning the magnetic field parallel instead of orthogonal to the momentum of an incoming neutron,
the magnetic interaction in the potential region or the free region of interest gives rise to the asymmetry between opposite helicity
components instead of ``spin-$z$" components (should be the spin-$x$ components in the coordinate frame defined here)
for the spin state of the scattered particle.
We observe that, when the barrier interaction dominates, such as the case of a particle tunneling through an opaque barrier, the helicity asymmetry is large, and correspondingly, $\tau^{T}_{~_\mathrm{Lar}}$ gets much closer to $\tau^{Tz}_{~L}$.
Whereas for the case of a transversely polarized particle reflected from an opaque barrier, or
transmitted through a free region or a low barrier with $A<E$ ($A$ is the maximum height of the barrier),
\ie, when magnetic interaction dominates,
the resultant helicity asymmetry is tiny and the spin evolution is largely confined in the transverse plane,
then our definition of $\tau^s_{~_\mathrm{Lar}}$ is dominated by $\tau^{sy}_{~L}$.

In fact, resembling the B$\mathrm{\ddot{u}}$ttiker-Landauer (BL) time
$\tau_{_\mathrm{BL}}\equiv\sqrt{(\tau_{~_\mathrm{LM}}^{y})^2+(\tau_{~_\mathrm{LM}}^{z})^2}$ \cite{Buttiker1983} defined as the
modulus of the complex time $\tau_{_\Omega}\equiv\tau_{~_\mathrm{LM}}^{y}-i\tau_{~_\mathrm{LM}}^{z}$ \cite{Sokolovski}\cite{LA1987},
where $\tau_{~_\mathrm{LM}}^y,~\tau_{~_\mathrm{LM}}^z$ are the conventional Larmor times,
we can also define the modulus $\tau_{_L}\equiv\sqrt{(\tau^{z}_{~L})^2+(\tau^{y}_{~L})^2}$.
Note we have omitted the superscripts $R$ and $T$ unless the distinction is necessary.
Interestingly, for the barriers we considered, $\tau_{_L}=\tau_{_\mathrm{Lar}}$, just as expected.
Also it is not surprising that for opaque barriers, $\tau_{_\mathrm{Lar}}$ gets close to $\tau_{_\mathrm{BL}}$,
and hence to the semi-classical time $\tau_{_\mathrm{SC}}\equiv\int_{_\Omega}\,dx\sqrt{m/[2(V(x)-E)]}$, since $\tau_{~_\mathrm{BL}}\simeq\tau_{_\mathrm{SC}}$ for opaque barriers.

For the linear barrier, we find
\bea\label{Larmory}
\tau_{~_\mathrm{L}}^{y}=\langle S_y\rangle/(-\frac{\hbar}{2}\omega_{_\mathrm{Lar}})
=-\frac{1}{\ga_{_N}B_0}\frac{2FG}{F^2+G^2}
\eea
where we have set $\al=0$ for calculational simplicity.
Also we have omitted the superscript $R$ in $\tau_{~_\mathrm{L}}^{y}$, 
since there is no transmitted flux for a linear barrier.
As mentioned above, for the reflective particle off a linear barrier, $\tau_{_\mathrm{Lar}}$ is dominated by $\tau_{~_\mathrm{L}}^{y}$,
thus $\tau_{~_\mathrm{L}}^{y}$ also gets closer to $\tau_{_\mathrm{CE}}$ with decreasing $B_0$ as $\tau_{_\mathrm{Lar}}$.
However, unlike the case of $\tau_{_\mathrm{Lar}}$, $\tau_{~_\mathrm{L}}^{y}$ monotonically deviates away from the classical returning time $\tau_{_\mathrm{CE}}$ with increasing $k$, rather than oscillating around it.

Another interesting observation for linear barrier is that, if we formally follow the standard definition
$\tau_{_\mathrm{LM}}\equiv-\hbar\frac{\prt\phi_R}{\prt\,V}$ \cite{Yamada2004}\cite{TTWM2016},
where $\phi_R$ is the phase of the reflective amplitude $\mathcal{R}\equiv|\mathcal{R}|e^{i\phi_R}$ and $V$ is the height of
the barrier, and define the reflective Larmor time as
\bea\label{RLMT}
\tau^R_{~_\mathrm{LM}}\equiv-\hbar\frac{\prt\phi_R}{\prt\,V_g}
=-2\hbar\left[\frac{\prt\theta_R}{\prt\,g}\frac{\prt\,g}{\prt\,V_g}+\frac{\prt\theta_R}{\prt\,m}\frac{\prt\,m}{\prt\,V_g}\right],
\eea
where $V_g=mgz$ and $\phi_R=2\theta_R$ is independent of $z$, see the statements above Eq.(\ref{TRphase}).
Clearly, $\tau^R_{~_\mathrm{LM}}$ is $z$-dependent. Interestingly, by appropriately choosing
$z=(\hbar\,k)^2/(m^2g)$ and substituting it into (\ref{RLMT}), a direct calculation shows that $\tau^R_{~_\mathrm{LM}}=\tau^R_{_\mathrm{phase}}$, where $\tau^R_{_\mathrm{phase}}$ is given by Eqn. (\ref{TRphase}).
We think this ``coincident equality" is due to the special choice of $z=2z_{_C}$ and a particular feature of linear barrier,
since for other semi-infinite barriers, such as the step barrier $V=A\,\theta(x)$, $\tau^R_{~_\mathrm{LM}}=\frac{\hbar\,k}{\kappa\,A}\neq\tau^R_{_\mathrm{phase}}=\frac{2m}{\hbar\,k\kappa}$,
where $\kappa\equiv\sqrt{2m(A-E)/\hbar^2}$.

\section{Summary}\label{Sumy}
In this paper, we utilize phase time, dwell time and Larmor time to calculate the time delays for a neutron scattering off
the linear gravitational potential.
As the Earth gravitational field is very gentle due to the extreme weakness of gravity,
the time scale for a vertically injected UCN climbing the potential is on the order of sub-millisecond,
which is indicated by the naive estimates in Table \ref{timeE} and the following calculations.
As far as we know, the tunneling particles in most time measurements include photon \cite{photonT} and electron \cite{AttoClock}\cite{ASAS2019},
whose relevant time scales are very short, say, femosecond. A fraction of experiments use atoms \cite{AtomT}\cite{BECTT},
and the time scales are $10\sim100$ microseconds.
So sub-millisecond shall be experimentally realizable, though UCN may be not easy to manipulate.
However, if the time measurement of UCN in the gravitational potential is experimentally feasible,
it may probe the quantum nature of scattering states of linear potential in the temporal domain,
in complementary to the spatial domain quantum test of the discrete turning heights of gravitational bound states \cite{Nesvizhevsky2002}.

By comparing these times, we obtain the relation between Wigner phase time and dwell time, shown in Eq.(\ref{RelaTPDT}).
In the end of last section, we also find that the conventionally defined reflective Larmor time, Eq.(\ref{RLMT})
coincides with the phase time for the linear barrier.
For our definition of Larmor time, it reduces to the classical returning time $\tau_{_\mathrm{CE}}=2v/g$
in the zero magnetic field limit.
Actually, dwell time in the barrier region, $\tau_{_\mathrm{DW}}[0,+\infty]$, also equals $\tau_{_\mathrm{CE}}$.
We think this may not be an accident, but rather a temporal manifestation of the weak equivalence principle (WEP).
By subtracting off the $\tau_{_\mathrm{CE}}$ from these times, we obtain the corresponding time delays.
All these time delays tend to be vanishingly small with increasing neutron wave number $k$,
and the amplitude of Larmor time delay decreases gently with increasing $k$, compared with other time delays,
see Fig.\ref{QFLTr}.
The excellent fit of the envelope of phase time delay, $\tau_\mathrm{coh}$, and the analysis of dwell time delay
(for \eg, see $\tau_{_\mathrm{C}}$ in Fig.\ref{DWtau}) are all manifestations of the self-interference
between incoming and reflective partial waves \cite{Fertig}.
To further reveal the self-interference effects, we plot the reflective self-interference delay $\tau_{_\mathrm{IF}}$
\cite{Winful2006} for linear barrier, step barrier and $\tanh$-like barrier in Fig.\ref{tauIF},
where the peculiar self-interference delay of linear barrier is demonstrated transparently.
This can be attributed to the very gentleness and particular shape of linear barrier.
This peculiarity may be one of the reasons that WEP holds true even in quantum domain \cite{EPQMDG},
where the advance decrease of the wave amplitude lower than the classical turning height cancels exactly with
the tunneling induced quantum lag in the classical forbidden region.
Actually, the classical forbidden region contributes a small but indispensable part for the dwell time to match
with the CRT in the large $k$ limit.

It is also interesting to note that the Larmor time defined here does have operational meanings since no need
for zero magnetic field limit. The magnetic field in our calculation is on the order of $0.1$mG, which is
not very stringent for current technologies. Further, the configuration of magnetic field, initial spin and momentum
is quite distinct from conventional definitions, and thus provides an alternative for experimental realization.

Further, the formalism can be directly applied to neutral atoms such as lithium.
Lithium is of comparable mass with neutron, say, $m_{^7\mathrm{Li}}\sim7m_n$, and
the combined Laser and the rf-induced forced evaporative coolings have already been able to reach
$T\simeq300$nK for $^7\mathrm{Li}$ \cite{LithiumBEC}\cite{LithiumLT}.
Since alkali atom is much easier to manipulate \cite{LithiumM} and the sub-millisecond timing accuracy
is not technically very stringent, ultracold $^7\mathrm{Li}$ ($^6\mathrm{Li}$) may be a good candidate
to probe the quantum temporal behavior in the Earth gravitational field.
For example, for a $^7\mathrm{Li}$ atom with effective $T\simeq823$nK (corresponds to $v\simeq54$mm/s
and classical turning height $\simeq148\mu$m), the dwell time delay is on the order of $20\sim35\mu$s.

At last, we note that a more rigourous and complete treatment is to start with a wave packet.
In that case, all the time delays and the relevant processes we have calculated before, such as finding the spin expectation value
of the reflective wave packet, have to be averaged over the weighting factor (or the distribution in $k$-space) $\rho(k)$ of the wave packet,
see Eqn.(\ref{WPacket}). That involves very tedious calculations, and will be left to future work.

\section{Acknowledgement}
Z. Xiao thanks X.X. Lu for his help.
This work is partially supported by National Science Foundation of China under grant No. 11120101004,
No. 11875127, No. 11605056, No. 11974108, and the Fundamental Research Funds for the Central Universities under No. 2020MS052.

\appendix
\section{Integral of Airy function}
The integral of Airy function used in the main context is
{\small
\bea\label{AiSI}&&
\int_{0}^{z_R}dz\mathrm{Ai}^2[z/L_c-E_{_D}]=L_c\left\{[E_{_D}\mathrm{Ai}^2+\mathrm{Ai}'^2]\right.\nn&&~~~~~~
\left.+(\frac{z_R}{L_c}-E_{_D})\mathrm{Ai}[\frac{z_R}{L_c}-E_{_D}]^2
-\mathrm{Ai}'[\frac{z_R}{L_c}-E_{_D}]^2\right\}.\nn
\eea
}

\section{Spin-precession angle}\label{Methods}
For a spin-$\hf$ particle carrying non-zero magnetic moment, the spin precession equation is
\bea\label{SpinPre}
\frac{dS^i}{dt}=\frac{i}{\hbar}[-\vec{\mu}\cdot\vec{B},S^i]
=\tilde{g}\mu_N(\vec{S}\times\vec{B})^i.
\eea
From (\ref{SpinPre})
we get the spin precession frequency,
$\omega_L=\tilde{g}\mu_N|\vec{B}|={g}\mu_NB_0$, suppose the magnetic field $|\vec{B}|=B_0$.

Now assume initially the spin is polarized along the polar and azimuthal angles $(\beta,\gamma)$, \ie, the initial spin state is
$\frac{1}{\sqrt{2}}\left(\cos[\beta/2]e^{-i\gamma},\sin[\beta/2]\right)^T$.
For later comparison, we can write it in the standard way up to a normalization constant
\bea\label{standSS}
|\beta, \gamma\rangle=\frac{1}{\sqrt{2}}\left(\begin{array}{c}
  \cot[\beta/2]e^{-i\gamma} \\
  1
\end{array}\right).
\eea
For example, for the neutron polarized in the antipodal direction, $\beta\rightarrow\beta+\pi$,
the spin state is
$
\frac{1}{\sqrt{2}}\left(-\tan[\beta/2]e^{-i\gamma}, 1\right)^T.
$
Assume the magnetic field is in the $z$-direction ($\theta=0$ in the main context)
and the initial state is polarized horizontally with the
azimuthal angle $\alpha$, in other words,
$$
|\pi/2, \alpha\rangle\langle\,z|k\rangle=\frac{1}{\sqrt{2}}\left(\begin{array}{c}
  e^{-i\alpha} \\
  1
\end{array}\right)e^{ikz}.
$$
The reflected partial wave is in the spin state
\bea\label{spinR}
|\pi/2+\delta\vartheta, \alpha+\delta\phi\rangle\langle\,z|-k\rangle
=\frac{1}{\sqrt{2}}\left(\begin{array}{c}
  \mathcal{R}_+ \\
  \mathcal{R}_-
\end{array}\right)e^{-ikz},
\eea
where $\delta\vartheta,~\delta\phi$ are the change of polarization angles.
From the standard form of spin state (\ref{standSS}), we can immediately read out
\bea
\cot[\hf(\pi/2+\delta\vartheta)]=|\frac{\mathcal{R}_+}{\mathcal{R}_-}|=1\Rightarrow\delta\vartheta=0,
\eea
so the spin evolves on the horizontal large circle on the Bloch sphere, which is dictated by the
spin evolution equation (\ref{SpinPre}). The rotated angle is then purely $\delta\phi$,
and can be determined by the following two methods.

\subsection{method A}
First note that
{\small\bea\label{rprn}&&
\frac{\mathcal{R}_+}{\mathcal{R}_-}=\nn&&
~e^{-i\al}\frac{(\mathrm{Ai}_+\mathrm{Ai}_-+\frac{1}{k_{_D}^2}\mathrm{Ai}'_+\mathrm{Ai}'_-)
+\frac{i}{k_{_D}}(\mathrm{Ai}'_+\mathrm{Ai}_--\mathrm{Ai}'_-\mathrm{Ai}_+)}
{(\mathrm{Ai}_+\mathrm{Ai}_-+\frac{1}{k_{_D}^2}\mathrm{Ai}'_+\mathrm{Ai}'_-)
-\frac{i}{k_{_D}}(\mathrm{Ai}'_+\mathrm{Ai}_--\mathrm{Ai}'_-\mathrm{Ai}_+)}\nn&&~~
=e^{-i(\al+\delta\phi)},
\eea }
where $-\delta\phi$ is the rotated angle of the spin on the horizontal large circle. 
From (\ref{rprn}), we find
\bea\label{spin1}&&
-\tan[\delta\phi/2]=\frac{k_{_D}(\mathrm{Ai}'_+\mathrm{Ai}_--\mathrm{Ai}'_-\mathrm{Ai}_+)}
{(k_{_D}^2\mathrm{Ai}_+\mathrm{Ai}_-+\mathrm{Ai}'_+\mathrm{Ai}'_-)}\Rightarrow\nn&&
\delta\phi=-2\tan^{-1}\left[\frac{k_{_D}(\mathrm{Ai}'_+\mathrm{Ai}_--\mathrm{Ai}'_-\mathrm{Ai}_+)}
{(k_{_D}^2\mathrm{Ai}_+\mathrm{Ai}_-+\mathrm{Ai}'_+\mathrm{Ai}'_-)}\right].
\eea
So the time measured by the Larmor clock is
{\small\bea
\tau_{\mathrm{Lar}}\equiv\frac{\delta\phi}{\omega_{\mathrm{Lar}}}=\frac{2m_{_I}}{k^2r\hbar}
\tan^{-1}\left[\frac{k_{_D}(\mathrm{Ai}'_+\mathrm{Ai}_--\mathrm{Ai}'_-\mathrm{Ai}_+)}
{(k_{_D}^2\mathrm{Ai}_+\mathrm{Ai}_-+\mathrm{Ai}'_+\mathrm{Ai}'_-)}\right],
\eea }

\subsection{method B}
The other method is to calculate the expectation value of spin vector $\hat{\vec{S}}=\frac{\hbar}{2}\vec{\si}$.
Substituting the reflective spin state (\ref{spinR}), we can get
{\small
\bea\label{Sz}&&
\langle S_z\rangle=\frac{1}{4}[|\mathcal{R}_+|^2-|\mathcal{R}_-|^2],\quad
\langle S_y\rangle=\frac{1}{4i}[\mathcal{R}_+^*\mathcal{R}_--c.c],\nn&&~~
\langle S_x\rangle=\frac{1}{4}[\mathcal{R}_+^*\mathcal{R}_-+c.c].
\eea
}
In the case $\theta=0$, where the magnetic field is along $z$-direction, direct calculation gives
$\langle S_z\rangle=0$, and
{\small
\bea\label{Syx}&&
\langle S_y\rangle=\frac{2GF\cos\alpha+(F+G)(F-G)\sin\alpha}
{2[\mathrm{Ai'}_-^2+(k_{_D}\mathrm{Ai}_-)^2]
[\mathrm{Ai'}_+^2+(k_{_D}\mathrm{Ai}_+)^2]},\nn&&
\langle S_x\rangle=\frac{(F+G)(F-G)\cos\alpha-2GF\sin\alpha}
{2[\mathrm{Ai'}_-^2+(k_{_D}\mathrm{Ai}_-)^2]
[\mathrm{Ai'}_+^2+(k_{_D}\mathrm{Ai}_+)^2]},
\eea}
where $F\equiv[\mathrm{Ai'}_+\mathrm{Ai'}_-+k_{_D}^2\mathrm{Ai}_-\mathrm{Ai}_+], G\equiv[\mathrm{Ai}_+\mathrm{Ai'}_--\mathrm{Ai}_-\mathrm{Ai'}_+]k_{_D}$, and the denominator in (\ref{Syx})
is $2(F^2+G^2)$. Note $\langle S_z\rangle=0$ means the reflected spin state is on the horizontal plane, \ie,
$\delta\vartheta=0$, so the azimuthal angle of the new spin state can be obtained from
\bea
\tan[\alpha+\delta\phi]=\frac{\langle S_y\rangle}{\langle S_x\rangle}=\frac{2GF/(F^2-G^2)+\tan\alpha}{1-2GF/(F^2-G^2)\tan\alpha}.
\eea
Finally we can read out the rotated angle
\bea\label{spin2}
\delta\phi=\arctan[2GF/(F^2-G^2)]=2\arctan[G/F].
\eea
It is easy to check that (\ref{spin2}) is exactly the same as (\ref{spin1}), confirming our calculations.

\section{Larmor time for free motion and square barrier}\label{FSLarmorT}
In this section, we discuss two special cases with our definition of Larmor time, the free motion and rectangular barrier.
For the former case, we will show that it indeed reduces to the classical traversal time as expected, while
for the latter case, the definition also generates reasonable results.
To show that our Larmor time definition doesn't rely on special coordinate configurations, different from the main
text, we choose instead the $x$-coordinate as the moving direction, while the particle is initially polarized in the spin-$z$ eigenstate.
\subsection{Free motion}
The relevant Hamiltonian for free particle is
\bea&&
\hat{H}_{\mathrm{Free}}=\frac{\hat{\vec{p}}^2}{2m_I}
-\frac{\hbar\,\mu_N\tilde{g}B_0}{2}\vec{\sigma}\cdot\hat{n}\Theta(x)\Theta(a-x),
\eea
where $[0,a]$ is the space interval we are interested, and is covered by a homogeneous weak magnetic field $\vec{B}=B_0\hat{n}$.
As mentioned in the main text, $\hat{n}$ is chosen to be parallel to the $x$-direction, \ie, $\hat{n}=\hat{x}$.
For simplicity, consider the stationary solution with a monochromatic neutron beam.
The ansatz of the corresponding wave function is
\bea\label{ansatz2}
\phi(x)=\left\{\begin{array}{c}
          \frac{1}{\sqrt{2}}\left[\left(
                     \begin{array}{c}
                       1 \\
                       0 \\
                     \end{array}
                   \right)e^{ikx}+ \left(
                     \begin{array}{c}
                       \mathcal{R}^f_+ \\
                       \mathcal{R}^f_- \\
                     \end{array}
                   \right)e^{-ikx}\right],~~~x<0,
          \\
          U\left(
             \begin{array}{c}
               c_+e^{ik_+\,x}+d_+e^{-ik_+\,x} \\
               c_-e^{ik_-\,x} +d_-e^{-ik_-\,x} \\
             \end{array}
           \right), ~~~~0<x<a,\\
              \frac{1}{\sqrt{2}}\left(
                     \begin{array}{c}
                       \mathcal{T}^f_+ \\
                       \mathcal{T}^f_- \\
                     \end{array}
                   \right)e^{ik\,x},~~~x>a,
          \end{array}\right.
\eea
where
$U\equiv\frac{1}{\sqrt{2}}\left(
                             \begin{array}{cc}
                               1 & 1 \\
                               1 & -1 \\
                             \end{array}
                           \right)
$, $k^2\equiv\frac{2mE}{\hbar^2}$ and $k_\pm^2\equiv2m(E\pm\frac{\hbar\,\mu_N\tilde{g}\,B_0}{2})/{\hbar^2}$.
Imposing the continuity conditions at $x=0$ and $x=a$,
we get the following solutions for the transmitted and reflective coefficients
\bea\label{Transm}&&
\mathcal{T}^f_+
=\hf[\mathcal{T}_{k_+}+\mathcal{T}_{k_-}],\quad  
\mathcal{T}^f_-
=\hf[\mathcal{T}_{k_+}-\mathcal{T}_{k_-}],\\&&\label{Reflec}
\mathcal{R}^f_+=\hf[\mathcal{R}_{k_+}+\mathcal{R}_{k_-}],~~
\mathcal{R}^f_-
=\hf[\mathcal{R}_{k_+}-\mathcal{R}_{k_-}].
\eea
where $\mathcal{T}_\rho\equiv\frac{2ik\rho\,e^{-ik\,a}}{(k^2+\rho^2)\sin(\rho\,a)+2ik\rho\cos(\rho\,a)}$
and $\mathcal{R}_\rho\equiv\frac{(k^2-\rho^2)\sin(\rho\,a)}{(k^2+\rho^2)\sin(\rho\,a)+2ik\rho\cos(\rho\,a)}$ are
the transmitted and reflective amplitudes for a rectangular barrier (or well), respectively,
and $\rho$ takes values of ${k_+},\, {k_-}$.
\begin{figure}
 \centering
 \subfigure[~Larmor time vs classical expectation time $\tau_{_\mathrm{CT}}$]{\includegraphics[width=72mm]{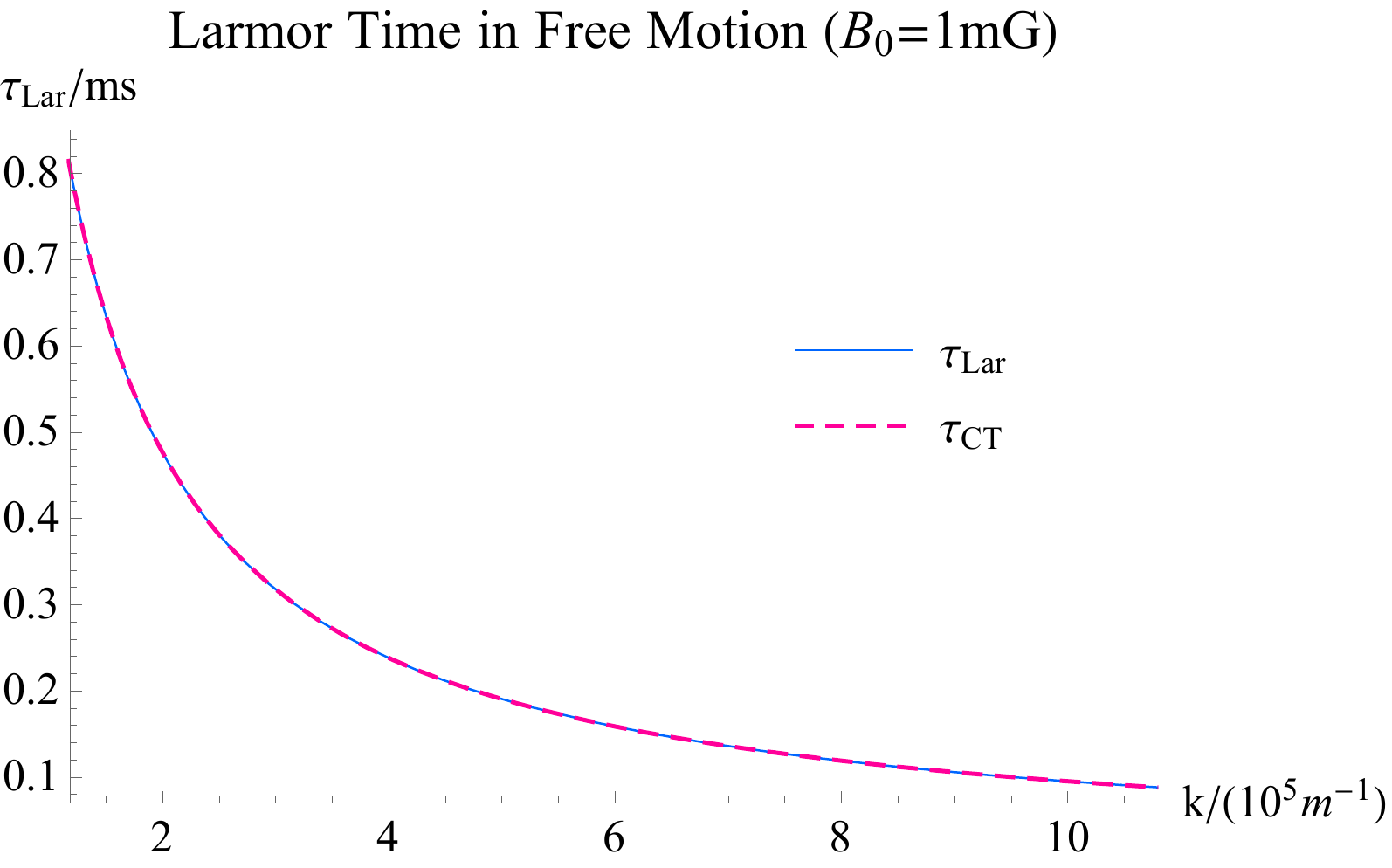}\label{FreeLar1}}
 \subfigure[~Relative time error $\de\tau$ with respect to $k$ and $B_0$ ]{\includegraphics[width=72mm]{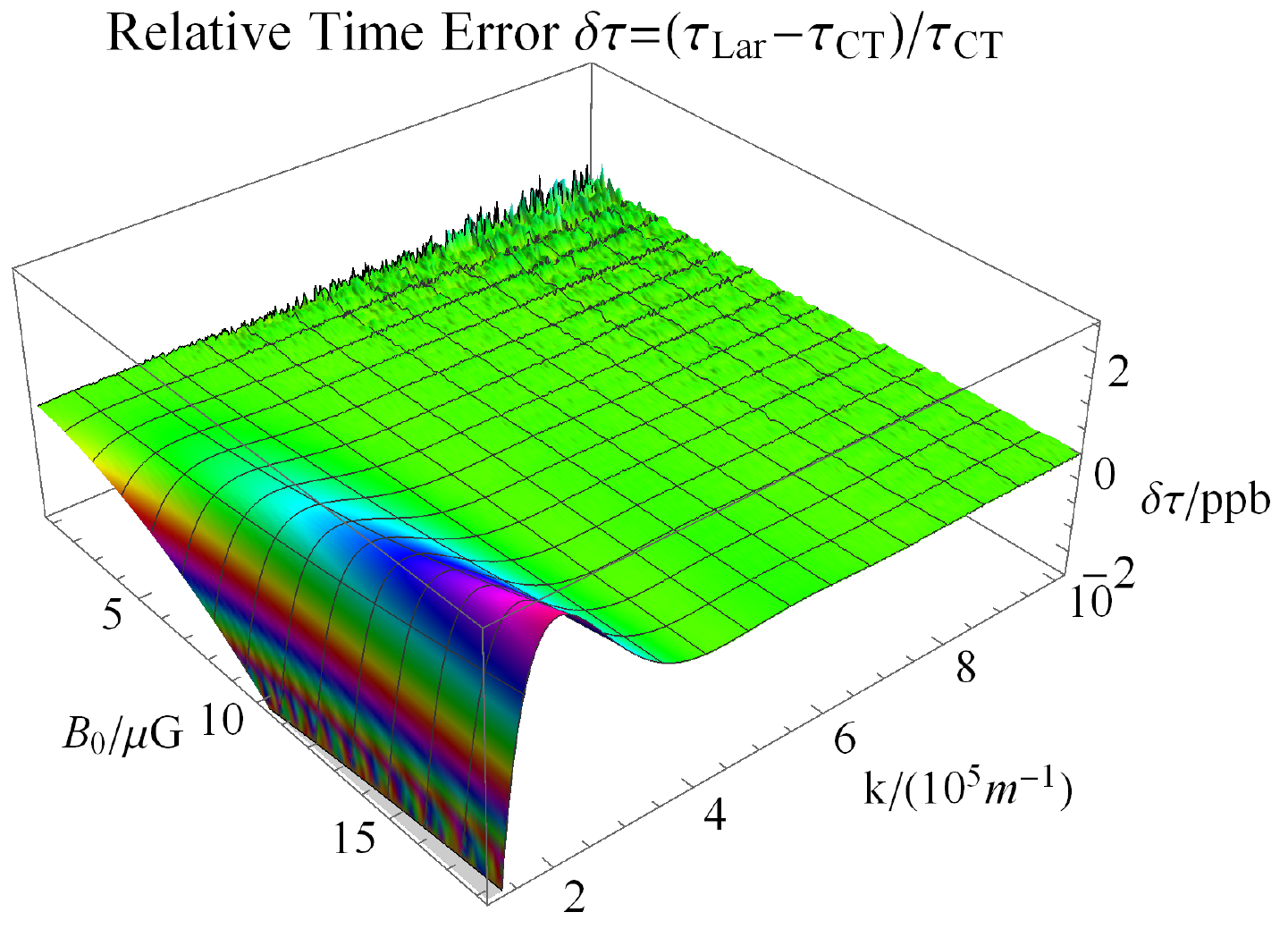}\label{RelError}}
     \caption{~a. Larmor time vs classical expectation time $\tau_{_\mathrm{CT}}$ with respect to the wave number $k$, represented by the solid blue and dashed red curves respectively. The space interval is $a=6\mu$m and the external magnetic field is $B_0=1$mG$=10^{-3}$G.
     b. Relative time error $\de\tau$ with respect to $k$ and $B_0$ in unit of part per billion (ppb). The length of the interested
     free interval is $a=10.56\mu$m. Here the external magnetic field is even smaller, in unit of $\mu$G=$10^{-6}$G.
     }\label{FreeT}
\end{figure}
The spin of the incoming particle is along the positive $z$-direction and is denoted as
$|\theta_i,\phi_i\rangle=|0,\phi_i\rangle=(1,0)^\mathrm{T}$,
while the spin for the transmitted particle is denoted as $|\theta_f,\phi_f\rangle$.
Since in the zero magnetic field limit, $\mathcal{T}^f_-\rightarrow0$, we'd better normalize the
spin state with the polar and azimuthal angles $(\beta,\gamma)$ by
\bea\label{standSS2}
|\beta, \gamma\rangle=\frac{1}{\sqrt{2}}\left(\begin{array}{c}
  1 \\
  \tan[\beta/2]e^{i\gamma}
\end{array}\right).
\eea
According to this normalization, the polarization of the transmitted partial wave can be written as
\bea\label{transmittS}
|\theta_f,\phi_f\rangle=\frac{1}{\sqrt{2}}\left(\begin{array}{c}
  1 \\
  \frac{\mathcal{T}^f_-}{\mathcal{T}^f_+}
\end{array}\right)=\frac{1}{\sqrt{2}}\left(\begin{array}{c}
  1 \\
  \frac{\mathcal{T}_{k_+}-\mathcal{T}_{k_-}}{\mathcal{T}_{k_+}+\mathcal{T}_{k_-}}
\end{array}\right).
\eea
Comparing (\ref{standSS2}) with (\ref{transmittS}), we can readily get
\bea
\tan[\theta_f/2]e^{i\phi_f}=\frac{\mathcal{T}_{k_+}-\mathcal{T}_{k_-}}{\mathcal{T}_{k_+}+\mathcal{T}_{k_-}}.
\eea
Since at the north pole $\phi_i$ can be arbitrary chosen, for convenience, we can chose $\phi_i=\phi_f$,
then we only need the modulus to read out the rotation angle
\bea
\theta_f=2\arctan[|\frac{\mathcal{T}_{k_+}-\mathcal{T}_{k_-}}{\mathcal{T}_{k_+}+\mathcal{T}_{k_-}}|],                                                                      \eea
and the Larmor time is given by
\bea\label{LarTF}
\tau_\mathrm{Lar}=\frac{\theta_f}{\omega_\mathrm{Lar}}=\frac{2}{\gamma_\mathrm{N}B_0}
\arctan[|\frac{\mathcal{T}_{k_+}-\mathcal{T}_{k_-}}{\mathcal{T}_{k_+}+\mathcal{T}_{k_-}}|],
\eea
where $\omega_{\mathrm{Lar}}\equiv\ga_{_N}B_0$ is the Larmor frequency and
$\ga_{_N}=1.832\times10^8\mathrm{rad}/(\mathrm{s}\mathrm{T})$ is the gyromagnetic ratio of neutron.
In Fig.\ref{FreeT}, we plot the Larmor time with respect to classical traversal time $\tau_{_\mathrm{CT}}\equiv{am}/(\hbar\,k)$
and the relative time error $\de\tau\equiv|\tau_\mathrm{Lar}-\tau_\mathrm{CT}|/\tau_\mathrm{CT}$.
From Fig.\ref{FreeT}, we see the Larmor time given by (\ref{LarTF}) fits well with the classical traversal time for two arbitrarily chosen space interval $a=6\mu$m and $a=10.56\mu$m.
In Fig.\ref{RelError}, we also see the relative time error grows large only around relatively small $k$ and large $B_0$.
This is as expected because the magnetic interaction becomes more relevant when $\frac{\hbar\,\mu_N\tilde{g}\,B_0}{2E}=\frac{m\mu_N\tilde{g}\,B_0}{\hbar\,k^2}$ grows large,
and this also drives the neutron away from free motion.

\begin{figure*}
 \centering
 \subfigure[~Reflective Larmor time vs free interval $a$ ahead of barrier]{\includegraphics[width=66mm]{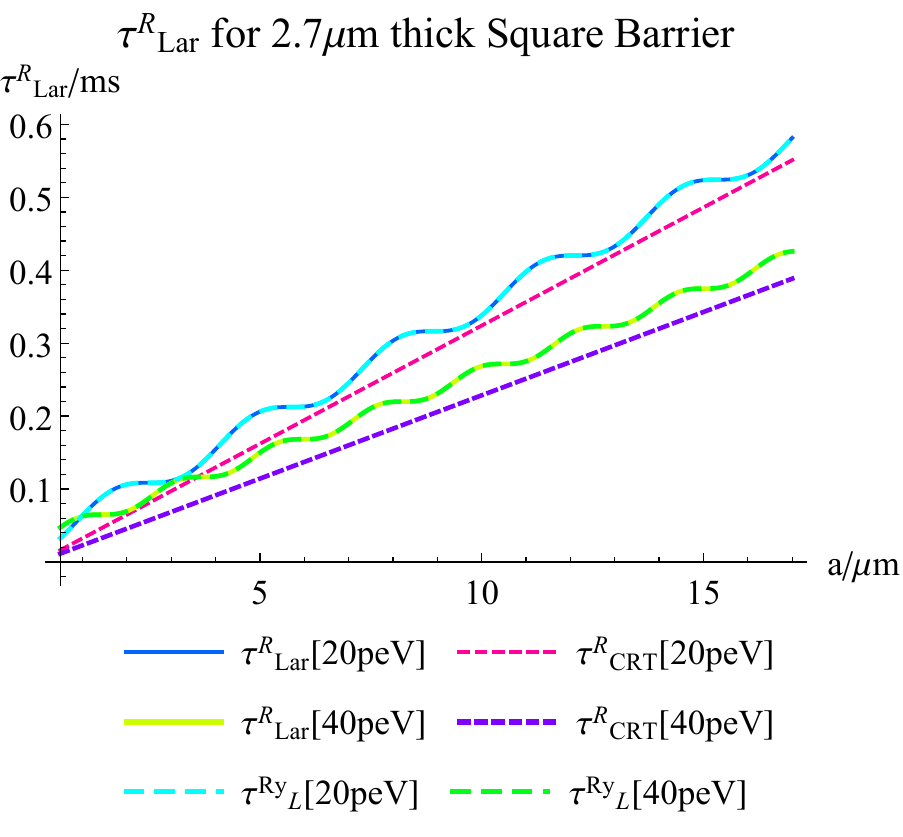}\label{LarTRSB1}}
  \hspace{0.2in}
 \subfigure[~Transmitted Larmor time vs barrier length $L$]{\includegraphics[width=66mm]{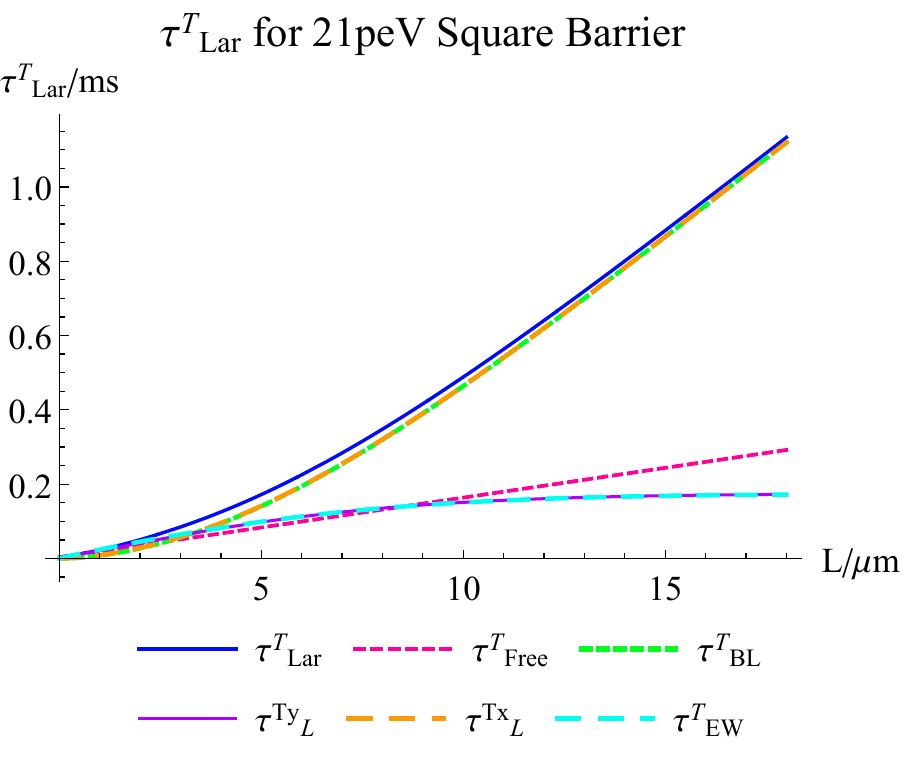}\label{LarTTSB1}}
     \caption{~a. Reflective Larmor time $\tau^R_{~~_\mathrm{Lar}}$ and classical returning time $\tau^R_{~~_\mathrm{CRT}}$ with respect to the free space interval $a$ in front of the barrier. For comparison, we also plot $y$-component Larmor time $\tau^{Ry}_{~~~L}$.
     These times are divided into two classes with different incoming neutron energies, which are labeled in the square bracket in the legends. The height and width of the barrier are $48$peV and $L=2.7\mu$m, respectively.
     b. Transmitted Larmor time $\tau^T_{~~\mathrm{Lar}}$ and B$\ddot{\mathrm{u}}$ttiker-Landauer (BL) time $\tau_{~~_\mathrm{BL}}^T$ with respect to the barrier width $L$. For comparison, we also plot component Larmor times $\tau^{Tx}_{~~~_L}$, $\tau^{Ty}_{~~~_L}$, free traversal time $\tau^T_{~~\mathrm{Free}}$ and Wigner phase time $\tau_{~~_\mathrm{EW}}^T$. The incoming neutron energy is $20.3$peV, the barrier height is $21$peV and the free space interval is $a=0.2\mu$m. In both subfigures, the external magnetic field is chosen as $B_0=0.1$mG.
     }\label{SBRTa}
\end{figure*}
\begin{figure*}
 \centering
 \subfigure[~k-spectrum of reflective Larmor time]{\includegraphics[width=66mm]{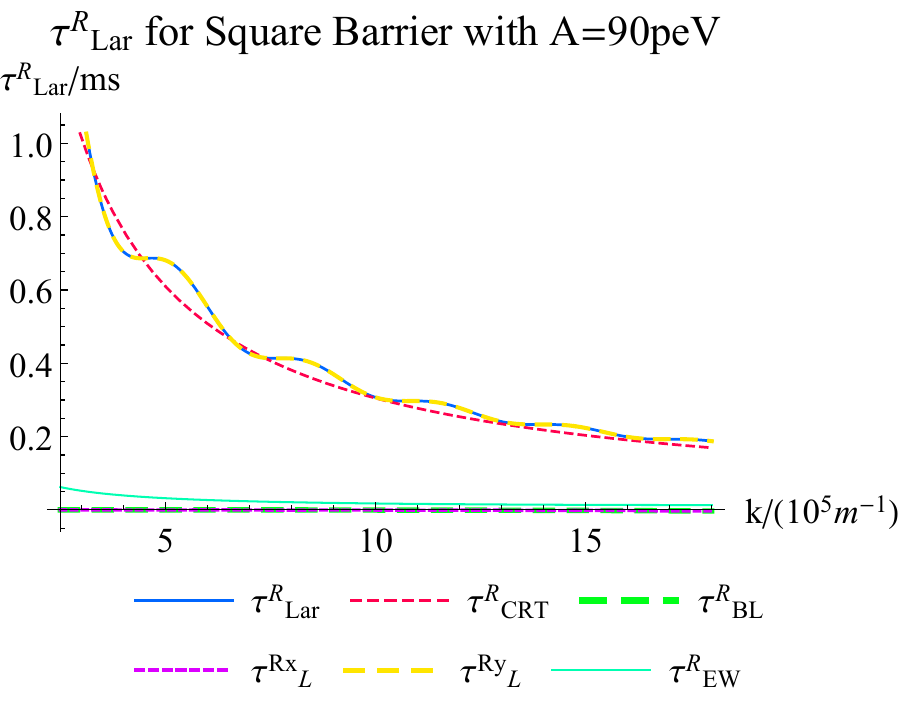}\label{LarTRSB2}}
  \hspace{0.2in}
 \subfigure[~k-spectrum of transmitted Larmor time]{\includegraphics[width=66mm]{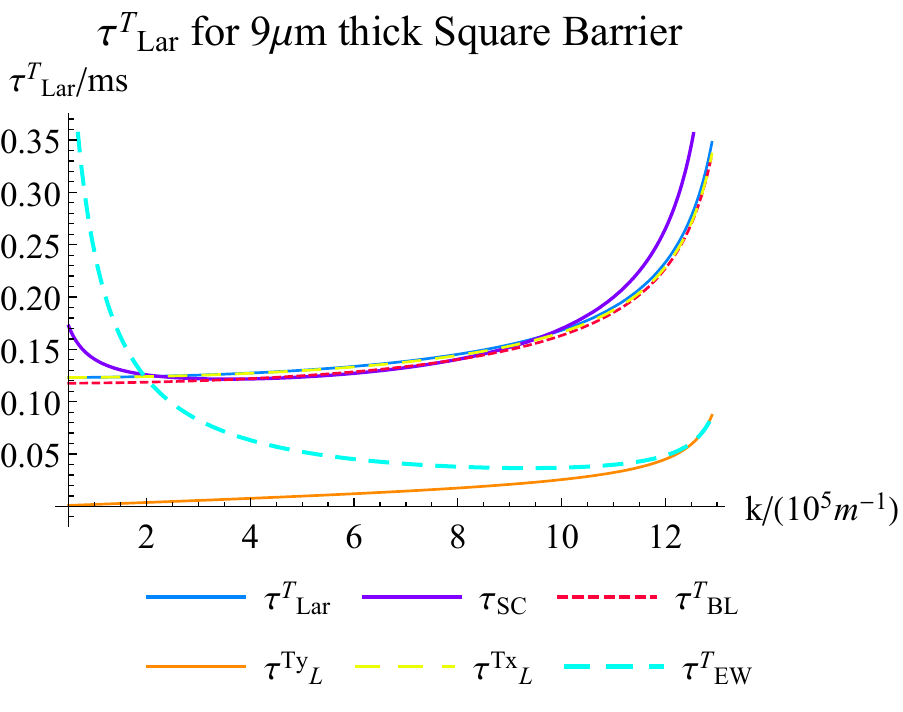}\label{LarTTSB2}}
     \caption{~a. The wave number $k$-spectrum of reflective Larmor times $\tau^R_{~_\mathrm{Lar}}$
     and the classical returning time. For comparison, we also show the Wigner phase time $\tau_{~~_\mathrm{EW}}^R$, BL time $\tau_{~~_\mathrm{BL}}^R$ and component Larmor times $\tau^{Rx}_{~~~_L},~\tau^{Ry}_{~~~_L}$ together
     in the subfigure. The barrier parameters are: height $A=90$peV and height $L=1.2\mu$m, and the free space interval $a=9.6\mu$m.
     b. The wave number $k$-spectrum of transmitted Larmor time $\tau^T_{~~\mathrm{Lar}}$ with transmitted BL time $\tau_{~~_\mathrm{BL}}^T$.
     In comparison, we also plot the component Larmor times $\tau^{Tx}_{~~~_L},~\tau^{Ty}_{~~~_L}$, together with semi-classical time $\tau_{_\mathrm{SC}}$ and Wigner phase time $\tau_{~~_\mathrm{EW}}^T$. The barrier parameters are $A=36$peV and $L=9\mu$m, and $a=0.2\mu$m. In both subfigures, the external magnetic field is chosen as $B_0=0.1$mG.
     }\label{SBRTk}
\end{figure*}
\subsection{Rectangular Barrier}
For a rectangular barrier, the relevant Hamiltonian in calculating Larmor time is
\bea\label{HSBLar}&&
\hat{H}_{\mathrm{SB}}=\frac{\hat{\vec{p}}^2}{2m_I}+A\,\Theta(x)\Theta(L-x)\nn&&~~~~
-\frac{\hbar\,\mu_N\tilde{g}B_0}{2}\vec{\sigma}\cdot\hat{n}\Theta(x+a)\Theta(L-x),
\eea
where $a>0$ is to allow for comparison of reflective larmor time with
classical returning time.
Though it seems more reliable and complete to start with a wave packet,
and in that case, for a wave packet with finite width $\de\,k$ in $k$-space,
it is suitable to choose $a>1/\de\,k$, we still work in the monochromatic limit,
since the wave packet formalism evolves integration over wave number $k$, and hence
is not easy or even impossible to obtain simple analytical formulas.
For the wave packet formalism, we leave it to future work.
Then the ansatz of the stationary wave function is
\bea\label{ansatz3}
\phi(x)=\left\{\begin{array}{c}
          \frac{1}{\sqrt{2}}\left[\left(
                     \begin{array}{c}
                       1 \\
                       0 \\
                     \end{array}
                   \right)e^{ikx}+ \left(
                     \begin{array}{c}
                       \mathcal{R}^s_+ \\
                       \mathcal{R}^s_- \\
                     \end{array}
                   \right)e^{-ikx}\right],~~~x<-a,
          \\
          U\left(
             \begin{array}{c}
               f_+e^{ik_+\,x}+g_+e^{-ik_+\,x} \\
               f_-e^{ik_-\,x} +g_-e^{-ik_-\,x} \\
             \end{array}
           \right), ~~~~-a<x<0,\\
          U\left(
             \begin{array}{c}
               c_+e^{\kappa_+\,x}+d_+e^{-\kappa_+\,x} \\
               c_+e^{\kappa_-\,x} +d_-e^{-\kappa_-\,x} \\
             \end{array}
           \right), ~~~~0<x<L,\\
              \frac{1}{\sqrt{2}}\left(
                     \begin{array}{c}
                       \mathcal{T}^s_+ \\
                       \mathcal{T}^s_- \\
                     \end{array}
                   \right)e^{ik\,x},~~~x>L,
          \end{array}\right.
\eea
where $k,\,k_\pm$ have already been defined in the subsection of free motion,
and $\kappa_\pm^2=2m(A-E\mp\frac{\hbar\,\mu_N\tilde{g}\,B_0}{2})/{\hbar^2}$. 
Imposing the continuity conditions at $x=-a,0,L$, we can get the reflective and transmitted coefficients as before.
Interestingly, these coefficients takes the similar form as in (\ref{Transm}), in other words,
\bea\label{TransmRB}
\mathcal{R}^s_\pm=\hf[\mathcal{R}^0_+\pm\mathcal{R}^0_-],\quad \mathcal{T}^s_\pm=\hf[\mathcal{T}^0_+\pm\mathcal{T}^0_-],
\eea
where $\mathcal{R}^0_\pm,~\mathcal{T}^0_\pm$ are just the reflective and transmitted amplitudes
for a spinless particle scattering off the barrier $A\,\Theta(x)\Theta(L-x)\mp\,V_0\Theta(x+a)\Theta(L-x)$,
where $V_0=\frac{\hbar\,\mu_N\tilde{g}B_0}{2}$. For example,
\bea&&
\mathcal{T}^0_{\rho}=\frac{e^{-ik(a+L)}}{\cos(k_\rho{a})A-i\sin(k_\rho{a})B},
\eea
where $A\equiv\ch(\kappa_\rho{L})+\frac{i}{2}(\frac{\kappa_\rho}{k}-\frac{k}{\kappa_\rho})\sh(\kappa_\rho{L})$
and $B\equiv\hf(\frac{k_\rho}{k}+\frac{k}{k_\rho})\ch(\kappa_\rho{L})
+\frac{i}{2}(\frac{\kappa_\rho}{k_\rho}-\frac{k_\rho}{\kappa_\rho})\sh(\kappa_\rho{L})$, and $\rho$ represents $\pm$.
Since the full expression is lengthy, we do not show explicitly the reflective amplitude here.
From the discussion of the free motion Larmor time, we can get the reflective and transmitted Larmor times as
\bea\label{RLarSB}&&
\tau_{~~_\mathrm{Lar}}^R=\frac{2}{\ga_NB_0}\tan^{-1}[|\frac{\mathcal{R}^s_-}{\mathcal{R}^s_+}|],\\&&
\label{TLarSB}
\tau_{~~_\mathrm{Lar}}^T=\frac{2}{\ga_NB_0}\tan^{-1}[|\frac{\mathcal{T}^s_-}{\mathcal{T}^s_+}|].
\eea
To show this definition is reasonable, we compare the reflective Larmor time $\tau_{~~_\mathrm{Lar}}^R$ with the classical returning time $\tau^R_{~~_\mathrm{CRT}}\equiv2am/(\hbar\,k)$, and also compare the transmitted Larmor time with
the Wigner phase time $\tau_{~~_\mathrm{EW}}^T=\prt\theta_\mathcal{T}/[v(k)\prt\,k]$ and
the B$\ddot{\mathrm{u}}$ttiker-Landauer (BL) time $\tau_{~~_\mathrm{BL}}^T=-\hbar\prt\ln|\mathcal{T}|/\prt\,A$.
To get an intuition, we plot them in Fig.\ref{SBRTa} and Fig.\ref{SBRTk}.
Further, we also plot the component Larmor times $\tau^{sx}_{~~~_L},~\tau^{sy}_{~~~_L}$ defined in (\ref{CLarYZ}),
only here the particle motion direction, \ie, the $z$-coordinate defined in the main context has been changed into $x$-coordinate.
In Fig.\ref{LarTRSB1}, the barrier height and width are $48$peV and $2.7\mu$m, respectively.
Comparing the distances of the solid blue curve to the dashed red line, and the solid yellow curve to the dashed purple line,
we find that the oscillating Larmor time $\tau_{~~_\mathrm{Lar}}^R$ gets more closer to the classical returning time $\tau_{_\mathrm{CRT}}$ for less energetic particles.
In other words, the more opaque the barrier is, the more it resembles a classical wall.
We can also see that $\tau^{Ry}_{~~~_L}$ nearly overlaps with $\tau_{~~_\mathrm{Lar}}^R$ from the two pairs curves.
This indicates that the spin precession for reflected partial wave is nearly confined in the transversal plane, \ie,
$y-z$ plane in this coordinates frame. This fact can also be confirmed from the nearly overlapped curves of $\tau^{Ry}_{~~~_L}$
and $\tau_{~~_\mathrm{Lar}}^R$ in Fig.\ref{LarTRSB2}, where $\tau^{Rx}_{~~~_L}$ is very close to the
BL time $\tau^R_{~\mathrm{BL}}$ and Wigner phase time $\tau^R_{~\mathrm{EW}}$,
but far smaller than $\tau_{~~_\mathrm{Lar}}^R$.
Note it is not easy to find $\tau^{Rx}_{~~~_L}$ and $\tau^R_{~\mathrm{BL}}$ there, since they nearly overlap with the horizontal $k$ axis.
We can also see in Fig.\ref{LarTRSB2} that, $\tau_{~~_\mathrm{Lar}}^R$ oscillates around the curve $\tau_{~~_\mathrm{CRT}}^R$,
which means that for thick barriers, disregard quantum fluctuations,
$\tau_{~~_\mathrm{Lar}}^R$ can give a good measure of the returning time.

For the transmitted Larmor time, we also plot the classical transversal time
$\tau_{_\mathrm{Free}}\equiv\frac{m(L+a)}{\hbar\,k}$ in Fig.\ref{LarTTSB1} as a comparison, see the dotted red line.
However, even for energetic particles (the particle's energy in Fig.\ref{LarTTSB1} is $20.3$peV compared to the barrier height $21$peV),
$\tau_{~~_\mathrm{Lar}}^T$ gets close to the free traversal time only for very thin barriers,
while for thick barrier, our transmitted Larmor time
resembles more closely to the BL time $\tau_{_\mathrm{BL}}^T$, as represented by
the dashed green curve.
The $\tau_{_\mathrm{SC}}\equiv\frac{ma}{\hbar\,k}+\frac{Lm}{\sqrt{2Am-(\hbar\,k)^2}}$ matches $\tau_{~~_\mathrm{Lar}}^T$
(the solid blue curve) only for particles with intermediate energy, see the solid purple curve.
We can also see that $\tau^{Tx}_{~~~_L}$ nearly coincides with $\tau_{_\mathrm{BL}}^T$, and is very close to
$\tau_{~~_\mathrm{Lar}}^T$, this can be further confirmed from Fig.\ref{LarTTSB2},
while $\tau^{Ty}_{~~~_L}$ is much smaller than $\tau_{~~_\mathrm{Lar}}^T$, and it is very close to Wigner phase time
$\tau_{~_\mathrm{EW}}^T$ for energetic neutrons, as can be also seen in Fig.\ref{LarTTSB2}.
So for opaque barriers, $\tau_{~~_\mathrm{Lar}}^T$ is dominated by $\tau_{~_\mathrm{Lx}}^T$.
This means that the helicity asymmetry caused by barrier interactions during tunneling is large, and
the spin precession cannot confined in the transversal plane. As a complementary observation,
the good match of $\tau_{~_\mathrm{EW}}^T$ with the sub-dominate $\tau_{~_\mathrm{Ly}}^T$ for energetic tunneling
particles confirms that Wigner phase time delay is not a meaningful measure of particle tunneling time.

In conclusion, we see our Larmor time definition is free of the coordinate choice and is rational,
as verified by the comparison of traversal time and the other frequently used time definitions
(such as the B$\ddot{\mathrm{u}}$ttiker-Landauer time) in free motion and square barrier cases.
A further clarification of the relation between the component Larmor times
$\tau_{~_\mathrm{Ly}}^s,~\tau_{~_\mathrm{Lx}}^s$ with our Larmor time definition $\tau_{~~_\mathrm{Lar}}^s$
and various discussed tunneling times in the literature will be interesting,
but is out of the scope of this work.

\end{document}